\documentclass[9pt]{livecoms}
\usepackage{lipsum} % Required to insert dummy text
\usepackage[version=4]{mhchem}
\usepackage{siunitx}
\usepackage{listings}
\usepackage{color}
\usepackage{pifont}
\usepackage{multirow}
\usepackage{dcolumn}
\usepackage{arydshln}
\usepackage{soul}
\newcolumntype{d}[1]{D{.}{.}{#1} }
 
\definecolor{codegreen}{rgb}{0,0.6,0}
\definecolor{codegray}{rgb}{0.5,0.5,0.5}
\definecolor{codepurple}{rgb}{0.58,0,0.82}
\definecolor{backcolour}{rgb}{0.95,0.95,0.95}
 
\lstdefinestyle{mystyle}{ 
    backgroundcolor=\color{backcolour},
    commentstyle=\color{codepurple},
    keywordstyle=\color{blue},
    numberstyle=\tiny\color{codegray},
    stringstyle=\color{codepurple},
    basicstyle=\footnotesize,
    breakatwhitespace=false,         
    breaklines=true,
    frame=single,
    captionpos=b,                    
    keepspaces=true,                 
    numbers=none,                    
    numbersep=5pt,                  
    showspaces=false,                
    showstringspaces=false,
    showtabs=false,                  
    tabsize=2,
    emph={vpcompressd,vmovups,vmulpd,vmovupd,vfmadd213pd,memset,memcpy,aligned,estimated,potential,speedup,vfmadd231pd,vmovaps,vbroadcastsd},
    emphstyle={\color{codegreen}},
    emph={[2]zmm0,zmm1,zmm2,zmm3,zmm4,zmm5,zmm6,zmm7,zmm8,zmm9,zmm16,zmm17,zmm18,zmm19,zmm20,unaligned},
    emphstyle={[2]\color{red}},
    emph={[3]xCORE,AVX512,qopt,zmm,usage,high,qopt,report,phase,report,5,S,mkl,sequential,qopenmp,stubs,&,falign, align, array64byte,functions},
    emphstyle={[3]\color{codepurple}},
    emph={[4]vector,cost,scalar,length,compress,vectorized,math,library,calls, **},
    emphstyle={[4]\color{blue}}
}
\DeclareSIUnit\Molar{M}
\usepackage[italic]{mathastext}
\graphicspath{{figures/}}
\newcommand{\lv}{\Large\verb}
\newlength{\mytabsav}
\usepackage{textcomp}
\newcommand{\rr}{\textsuperscript{\textregistered}}
\newcommand{\tm}{\textsuperscript{\texttrademark}}
\title{Raising the Performance of the Tinker-HP Molecular Modeling Package [Article v1.0]}

\author[1*]{Luc-Henri Jolly}
\author[2]{Alejandro Duran}
\author[1, 3]{Louis Lagardère}
\author[4]{Jay W. Ponder}
\author[5]{Pengyu Ren}
\author[6, 7, 5*]{Jean-Philip Piquemal}
\affil[1]{Institut Parisien de Chimie Physique et Théorique, Sorbonne Université, FR 2622 CNRS, 75005, Paris France.}
\affil[2]{Intel Corporation Iberia, Spain}
\affil[3]{Institut des Sciences du Calcul et des Données, Sorbonne Université, 75005, Paris France. }
\affil[4]{Department of Chemistry, Washington University in Saint Louis, Saint Louis, Missouri 63130, United States}
\affil[5]{Department of Biomedical Engineering, The University of Texas at Austin, Austin, Texas 78712, United States.}
\affil[6]{Laboratoire de Chimie Théorique, Sorbonne Université, UMR 7616 CNRS, 75005, Paris France.}
\affil[7]{Institut Universitaire de France, 75005, Paris, France}

\corr{luc-henri.jolly@sorbonne-universite.fr}{LHJ}
\corr{jean-philip.piquemal@sorbonne-universite.fr}{JPP}% Correspondence emails.  FMS and FS are the appropriate authors initials.

\blurb{}

\pubDOI{10.XXXX/YYYYYYY}
\pubvolume{<volume>}
\pubissue{<issue>}
\pubyear{<year>}
\articlenum{<number>}

\begin{document}

\begin{frontmatter}
\maketitle

\begin{abstract}
This living paper reviews the present High Performance Computing (HPC) capabilities of the Tinker-HP molecular modeling package. We focus here on the reference, double precision, massively parallel molecular dynamics engine present in Tinker-HP and dedicated to perform large scale simulations. We show how it can be adapted to recent Intel\rr Central Processing Unit (CPU) petascale architectures. First, we discuss the new set of Intel\rr Advanced Vector Extensions 512 (Intel AVX-512) instructions  present in recent Intel processors (e.g., the Intel\rr Xeon\rr Scalable and Intel\rr Xeon Phi\tm 2nd generation processors) allowing for larger vectorization enhancements.  These instructions constitute the central source of potential computational gains when using the latest processors, justifying important vectorization efforts for developers.  We then briefly review the organization of the Tinker-HP code and identify the computational hotspots which require Intel AVX-512 optimization and we propose a general and optimal strategy to vectorize those particular parts of the code. We present our optimization strategy in a pedagogical way so it can benefit other researchers interested in improving performances of their own software. Finally we compare the performance enhancements obtained to unoptimized code, both sequentially and at the scaling limit in parallel for classical non-polarizable (CHARMM) and polarizable force fields (AMOEBA). This paper will be updated as we accumulate new data available on the associated Github repository between versions of this living document.
\end{abstract}

\end{frontmatter}

\newpage
\hbox{} 

\newpage

\section{Introduction}

\hspace{\parindent}Tinker-HP is a massively MPI parallel package dedicated to classical molecular dynamics (MD) and to multiscale simulations, especially using advanced polarizable force fields (PFF) encompassing distributed multipoles electrostatics\cite{Tinker-HP}.
It is an evolution of the popular Tinker package code \cite{Tinker8} which conserves Tinker's simplicity of use and its developer-friendliness, allowing for the rapid development of new algorithms. Tinker-HP offers the possibility to perform large scale simulations while retaining the Tinker reference double precision implementation dedicated to CPU-based (Central Processing Unit) petascale architectures. The parallel scalability of the software is demonstrated via benchmark calculations. Overall, a several thousand-fold acceleration over a single-core computation is observed for the largest molecular systems, thus allowing long reference polarizable MD simulations on large molecular systems containing up to millions of atoms. 

Despite this strong acceleration, and due to the development model of the Tinker software suite (now version 8 \cite{Tinker8}), which favours new scientific development over optimization, no attempt has previoustly been made to adapt the Tinker-HP code to a particular CPU architecture. Each execution of the existing code took little or no advantage of the underlying capabilities of the CPU it was running on. The deployment of numerous processors with SIMD (Single Instruction/Multiple Data) capabilities that in principle allow substantial speedups leads us to address this issue. The existence of strong vector capabilities on modern Intel\rr  architectures, and particularly the Intel\rr Advanced Vector Extensions 512 (Intel AVX-512) on Intel\rr\ Xeon\rr\ Scalable and Intel Xeon\rr\ Phi\tm\ processors, combined with the excellent support from the Intel\rr\ Fortran Compiler, motivates us to change the overall design of the code, while trying to keep its simplicity and readability. The goal of this paper is two-fold. First, it provides a comprehensive living review dedicated to the capabilities and performance of Tinker-HP's main production methods (i.e. force fields) on Intel's architectures. Second, it is organized to present in a pedagogical fashion our code optimization efforts, outlining an optimal strategy for vectorization of Tinker-HP. Although there have been theoretical discussions of AVX-512 vectorization\cite{Shabanov2019}, and a description of the effective vectorization of a Quantum Monte-Carlo production code\cite{QMCpackavx}, such practical examples are rarely documented and we think our experience could prove useful to other software developers.

The present version of the paper is organized as follows. After reviewing the specificities of the latest Intel Xeon Scalable Processors (code-named Skylake) and particularly their Intel AVX-512 vector instruction set, we present the general structure of the most CPU intensive Fortran subroutines in Tinker-HP, show their performance hotspots and propose a general strategy to vectorize the code within the Intel AVX-512 instruction set. We then give concrete examples of the vectorization process. Performance comparisons between the current Tinker-HP release and the vectorized version is then made, first for execution on a single core, to show brute force acceleration of the code, and then in the context of a realistic parallel execution (with up to 16~000 cores). Finally, extended benchmarks on meaningful systems are provided with the AMOEBA polarizable force field \cite{ren2003polarizable,shi2013polarizable,zhang2018amoeba} and also using an initial implementation of classical force fields such as CHARMM \cite{Charmm}, to illustrate what can be achieved with a typical non-polarizable force field (AMBER,\cite{Amber} OPLS-AA\cite{jorgensen1996development} etc...).
\section{Intel Xeon Scalable processors}
\hspace{\parindent}We are using in our study a system with a CPU from the Intel Xeon Scalable processor family (code-named Skylake). These processors feature up to 28 cores per processor with two hyperthreads per core for a total of up to 56 threads per processor. A new mesh interconnect reduces the latency of communication within cores and controllers in the processor. Each core is capable of issuing up to four instructions per cycle, out-of-order. Up to two of these can be Intel AVX-512 instructions\cite{intel-avx512}. The Intel AVX-512 instruction set is a 512-bit SIMD instruction set that allows performing computations with a single instruction using SIMD registers that contain eight double-precision (DP) or sixteen single-precision (SP) floating-point values, as well as a variety of integer data sizes. The Intel AVX-512 instruction set also supports unaligned loads, fused-multiply and add, vector masking, shuffle and permutation instructions, histogram support, and hardware-accelerated transcendentals. Using all available cores and SIMD units is key to unlocking the full potential performance of these processors.

A significant change from its predecessor, the Intel Xeon processor v4, is the reorganization of the cache hierarchy to better balance the cache usage for server workloads. To achieve this, the L2 cache has increased in size to 1MB. The last level cache (LLC) has been reduced in size (up to 38.5 MBs), but it is now a non-inclusive cache (meaning that data is not evicted from caches closer to the cores when evicted from the LLC).

The Intel Xeon Scalable processors provide two memory controllers with three memory channels each that support DDR4 memory at up to 2600 MHz. This provides up to 123 GB/s of bandwidth to main memory for each socket. Three Intel\rr Ultra Path Interconnect links, each providing 10.4 GT/s, allow multiple processors to communicate to create bigger systems (e.g, dual-socket systems).

\section{Considerations on vectorization}
\label{subsection:Considerations_vecto}
\hspace{\parindent}Efficient use of the SIMD units available in the processor is very important to achieve the best performance on modern (and probably future) processors. Most vector parallelism is extracted from loops. In this section, we outline some ideas that can help to achieve good performance from loop vectorization. 

Modern compilers are able to auto-vectorize loops but need to be able to determine that the vectorization does not break any possible cross-iteration dependencies. This is not always possible due, for example, to variable aliasing or loop complexity\cite{vector-compiler}. Programmers can assist the compilers by rewriting their loops with idioms that compilers recognize or by using directives that guide the vectorization process.

Once a loop is vectorized, it is useful to consider if vector code generated for the loop is the most efficient possible. There are tools that can help with this assessment as well as provide feedback on how to improve efficiency\cite{vector-advisor}. Common issues that are worth looking into are:
\begin{itemize}
    \item \textbf{Unaligned loads or stores}. When data is not aligned to cache boundaries, some penalty will be incurred in load and store instructions compared to well aligned data. Therefore, it is recommended to align data to cache boundaries and also use the proper mechanisms to inform the compiler.
    \item \textbf{Loop prologues and remainders}. To provide better data alignment, the compiler might generate different code for the first iterations of the loop to ensure the main vectorized portion of the loop runs on aligned data. Similarly, if the compiler cannot deduce the number of loop iterations, it will generate a remainder loop to compute the final iterations of a loop that do not fill a vector register completely.   
    \item \textbf{Unnecessary masking}. While the Intel AVX-512 instruction set supports vectorization of loops with conditional statements, this feature requires the use of mask instructions and masked vector operations which reduces the vectorization efficiency. Therefore it is recommended to move, as much as possible, conditional statements out of vectorized loops. This might require spliting the loop (with different code for those iterations where the branch was taken vs. was not taken) or duplicating the loop (with different code for each branch of the conditional).
    \item \textbf{Non-unit strides}. Use of non-unit strides will often force the compiler to generate gather and/or scatter instructions which are less efficient than regular loads and stores. This also includes accessing a field in a structure, as a non-unit stride will be needed to access the same field between consecutive elements of an array. This is why a Struct-of-Arrays (SoA) layout is preferred over a more conventional Array-of-Structs (AoS) layout for efficient vectorization. Additionally, the Array-of-Structs-of-Arrays (AoSoA) layout can be used where a given number of elements of a field are placed consecutively in memory like in SoA, usually a small multiple of the cache line size, but element fields are still interleaved in memory as in AoS. This layout can improve locality when multiple fields are used closely together in a code section at the expense of increasing code complexity.
    \item \textbf{Indirect accesses}. Indexing one array with the values from another array will also result in generation of gather and/or scatter instructions. Therefore, this pattern should be avoided as much as possible.
    \item \textbf{Register pressure}. The number of vector registers is limited (e.g., Intel AVX-512 provides 32 vector registers). If the number of arrays that are used in a given loop, plus temporary values that might be required for the operation of the loop, exceeds the number of registers, then the compiler will need to spill some of the arrays to the stack and restore them afterwards which significantly reduces vectorization efficiency. To avoid this situation, it is better to use different loops for independent array operations rather a single big loop containing all arrays.
\end{itemize}

\section{Working environment and definitions}
\subsection{Definitions} 
\hspace{\parindent} In this paper, we will use two versions of Tinker-HP~: 
\begin{enumerate}
    \item the Release Version 1.1, referred to as \textbf{Rel}.
    \item the Vectorized Version 1.1v, referred to as \textbf{Vec}.
\end{enumerate}

A third version, Release Version 1.2 (referred to as \textbf{Rel2}), is mentioned in the Perspective subsection~\ref{subsection:Perspectives} to give a feeling for the performance gains anticipated with newly-developed algorithms. Vectorization of \textbf{Rel2} is in progress. It will be referred to as \textbf{Vec2}. 

We ran Tinker-HP exclusively on \emph{supercomputers} under \textsc{Unix/Linux} Operating System (OS). These machines aggregate hundreds or even thousands of interconnected systems called \emph{computing nodes}, or simply \emph{nodes}, each typically having tens of CPU cores and hundreds of gigabytes of memory. On \textsc{Unix/Linux} computers, the code is executed by a \emph{process}, which uses memory and CPU resources managed by the OS.

What we called \emph{the code} can be split in two parts~:
\begin{enumerate}
\item the \textbf{User Code (UC)}, which comprises all the Fortran code. Here, it is the code for Tinker-HP and the \textsc{2Decomp} library.
\item the \textbf{Non--User Code (NUC)}, which comprises all the code executed by the process because of library calls from the \textbf{UC}, system calls done on behalf of the \textbf{UC} or code introduced implicitly by the compiler.
\end{enumerate}

Of course, the way we write the \textbf{UC} (what library we used, how we setup our data,...) has an influence on the \textbf{NUC} executed by the process. As we want to raise the performance, we have to take into account what \textbf{NUC} gets executed because of the \textbf{UC} code that we write.

We use the term \emph{Molecular System} (\textbf{MS}) to denote all the physical systems we have simulated for this paper.

Note that Fortran code listings shown in this paper have been taken \emph{as is}, while all compilation reports and assembly code listings have been edited for publication purposes.
\subsection{Compilation setup}
\hspace{\parindent} We worked with the Intel\rr\ Parallel Studio XE 2018 development suite\cite{intel-parallel-studio}, containing the Intel Fortran Compiler, the Intel\rr\ MPI Library, the Intel\rr\ Math Kernel Library (Intel MKL) with implementations of \textsc{Blas}\cite{blas}, \textsc{Lapack}\cite{lapack} and \textsc{Fftw3}\cite{fftw3} routines, and the Intel\rr VTune\tm  Amplifier for profiling and analysis.

The Tinker-HP sources are compiled with the flags shown in listing~\ref{listing:Compil-flags} where~:
\begin{itemize}
    \item {\color{codepurple}\lv|-xCORE-AVX512|} flag forces the generation of binaries for the Intel Xeon Scalable processors.
    \item {\color{codepurple}\lv|-qopt-zmm-usage=high|} flag instructs the compiler to use {\color{red}zmm} (512 bits) registers as much as possible.
    \item {\color{codepurple}\lv|-align array64byte|} instructs the compiler to align all static arrays to 64 bits memory address boundaries
    \item {\color{codepurple}\lv|-falign-functions=64|} tells the compiler to align functions on 64 bits boundaries
    \item {\color{codepurple}\lv|-qopt-report-phase|} and {\color{codepurple}\lv|-qopt-report=5|} flags produce vectorization reports.
    \item {\color{codepurple}\lv|-S|} flag produces assembly code listings.
\end{itemize}
\begin{lstlisting}[label=listing:Compil-flags,language=make, firstline=1, lastline=5,caption=Flags used for the compilation of Tinker-HP with Intel Fortran compiler. ]
FFLAGS = -O3 -xCORE-AVX512 -qopt-zmm-usage=high
         -no-ipo -no-prec-div -shared-intel
         -align array64byte -falign-functions=64
         -qopt-report-phase=vec -qopt-report=5 -S 
         -qoverride-limits  
\end{lstlisting}

The vectorization reports are of major interest, because they give precise indications on how the compiler understands the code and what decisions it makes, allowing the programmer to modify the code or give the compiler pertinent directives. Some indications are of particular interest:

\begin{itemize}
    \item the \mbox{\em vector length}, which gives the number of elements (real, integer or logical) affected by one operation
    \item the \mbox{\em scalar cost}, which gives the number of scalar operations necessary to handle one iteration of the loop
    \item the \mbox{\em vector cost}, which gives the same information as the scalar cost, but for vector operations
    \item the \mbox{\em estimated potential speedup}. Most of the time, this is the ratio scalar cost/vector cost. This speedup is not in time of execution, but in number of operations. A speedup of 10.0 does not mean that a loop will execute 10 times faster, but rather that there will be 10 times fewer vector operations.
    \end{itemize}
 
Even if we do not use OpenMP*\cite{openmp} in the dynamic simulation engine, other parts of the Tinkertools suite (Tinker 8 \cite{Tinker8})  use OpenMP directives. So, object files are linked with the flags shown in listing~\ref{listing:Link-flags} where~:
\begin{itemize}
    \item {\color{codepurple}\lv|-mkl=sequential|} flag tells the linker to use the sequential Intel MKL library, which is lighter and faster than the multi-threaded ones (e.g., OpenMP).
    \item {\color{codepurple}\lv|-qopenmp-stubs|} flag enables compilation of OpenMP programs in sequential mode. 
\end{itemize}
\begin{lstlisting}[label=listing:Link-flags,language=make, firstline=6, caption=Flags used for the objects linking with Intel Fortran compiler.]
FFLAGS2 = -mkl=sequential -qopenmp-stubs
\end{lstlisting}

\subsection{Execution setup}
\hspace{\parindent}For the performance tests, calculations were performed on nodes running under the RedHat* Enterprise Linux* Server Release 7.4 operating system, with the following configuration~:
\begin{itemize}
    \item 2 $\times$ Intel Scalable Xeon 8168 processor -- 2,7 GHz -- 24 cores/processor
    \item 192 GB of DDR4 memory,
    \item InfiniBand* EDR interconnect.
\end{itemize}

We chose 8 \textbf{MS} from those studied in~\cite{Tinker-HP} with increasing sizes ranging from 9~737 to 3~484~755 atoms : the Ubiquitin protein, the prototypic Dihydrofolate Reductase (DHFR), the COX-2 dimer, the Satellite Tobacco Mosaic Virus (STMV), the Ribosome full structure in polarizable water, and three water boxes (Puddle, Pond and Lake). 

The table~\ref{table:MS} gives for each \textbf{MS} the name, the number of atoms and the number of cores used for the parallel calculations.
\begin{table}[htbp!]
\begin{tabular}{|l|d{0}|d{0}|d{0}|d{0}|}
  \hline
   \textbf{MS} & \multicolumn{1}{c|}{Ubiquitin}
           & \multicolumn{1}{c|}{DHFR} 
           & \multicolumn{1}{c|}{Puddle}
           & \multicolumn{1}{c|}{COX-2} \\
    \hline
   Atoms   &9~737 & 23~558 &96~000&174~219\\ 
  \hline
  CPU      & 480&680 &1~440 & 2~400  \\
  CPU2     &    &960 &      & 3~000 \\         
  \hline
  \hline
  \textbf{MS}   & \multicolumn{1}{c|}{Pond} 
                & \multicolumn{1}{c|}{Lake}
                & \multicolumn{1}{c|}{STMV} 
                & \multicolumn{1}{c|}{Ribosome}\\
  \hline
  Atoms    &288~000 & 864~000 &1~066~628 & 3~484~755\\
  \hline
  CPU      &2~400   &7~200    &10~800    &10~800\\
  CPU2     &        &         &16~200    &16~200 \\
  \hline
\end{tabular}
\caption{\textbf{MS} used for the performance measurements. The numbers of cores are taken from~\cite{Tinker-HP} for comparison. The CPU2 raw gives the number of cores which produced the best performance (See tables~\ref{table:timings_multi}, \ref{table:timings_nopol} and \ref{table:timings_multi2}). For the sequential performance measures, only one core was used.}
\label{table:MS}
\end{table}

All calculations have been made starting from an equilibrated \textbf{MS} with a timestep of $2 fs$ and a RESPA integrator\cite{tuckerman1992reversible}. For the one core tests, we define  the time $T$ (in \emph{seconds}) as the time of execution. For the parallel tests, we define the performance $P$ (in \emph{ns/day}) as the duration of simulation that can be achieved in one day. Our goal is to optimize throughput via vectorization. That means lowering the time $T$ and increasing the performance $P$. Thus, we define the boost factors $B$ as :
\begin{equation*}
    B=\frac{T_\mathbf{Rel}}{T_\mathbf{Vec}}
\end{equation*} 
where $T_\mathbf{Rel}$ and $T_\mathbf{Vec}$ are times for \textbf{Rel} and \textbf{Vec} respectively, or
\begin{equation*}
    B=\frac{P_\mathbf{Vec}}{P_\mathbf{Rel}}
\end{equation*} 
where $P_\mathbf{Rel}$ and $P_\mathbf{Vec}$ are performances for \textbf{Rel} and \textbf{Vec} respectively.

To get the profiles for \textbf{Rel} and \textbf{Vec}, we used DHFR \textbf{MS} with the AMOEBA polarizable force field and with the CHARMM classical force field (no polarization). Both simulations ran on one core and for 100 steps.
\section{Release version 1.1 of Tinker-HP}
\subsection{Polarizable AMOEBA force field}
\hspace{\parindent}We focus here on the part of the code dedicated to the AMOEBA polarizable force field which is the most computationally challenging and gives a lower bound for Tinker-HP performance\cite{ren2003polarizable,Ponder2007CurrentField}.
AMOEBA has been shown to have wide applicability for physical systems ranging from liquids to metals ions,\cite{Amber9AMOEBA,wu2010polarizable} including heavy metals,\cite{Marjolin2012,Marjolin2014} in gas and solution phase, to proteins \cite{shi2013polarizable,zhang2018amoeba} and to DNA/RNA\cite{zhang2018amoeba}. It uses distributed atomic multipoles through quadrupole moments and a Thole damped point dipole polarization interaction model. Van der Waals interactions use the Halgren buffered 14–7 function. In this paper, we used the AMOEBA 2013 protein parametrization \cite{shi2013polarizable,zhang2018amoeba} coupled with the 2003 water model\cite{ren2003polarizable}.
\subsection{General Structure}
\hspace{\parindent}Tinker-HP uses a 3D spatial decomposition to distribute atoms on compute cores. Every process is assigned to a subsection of the simulation box and is responsible of updating the positions, velocities and accelerations of the atoms present in this subdomain at each timestep\cite{Tinker-HP}. The most computationally intensive part of Tinker-HP is devoted to forces and electric fields calculations. 

All the compute routines follow the same organizational scheme~:
\begin{itemize}
    \item an external loop over all the atoms held by the process
    \item the selection of the neighbour sites using various criteria (e.g. cutoff distances, ...)
    \item a long internal loop over the selected sites, where all quantities are computed.
\end{itemize}

In \textbf{Rel}, this internal loop computes all quantities for each atom-neighbour pair \emph{on the fly}, with no attempt to pre-calculate or store intermediate quantities that can eventually be re-used. This results in a serious issue regarding cache registers and processing units, and in an extensive use of memory-core transfer instructions. By contrast, there's almost no use of arrays, besides indexing. This means data are often not contiguous in memory, and therefore accesses to memory are irregular. Thus, the possibility to take advantage of the vector extension capabilities of the Intel AVX-512 instructions is very low. 
\subsection{Hotspots}
\hspace{\parindent}The table~\ref{table:ReleaseProfUbiDHFR} shows the profiling of \textbf{Rel}, running on one core for DHFR with AMOEBA (polarizable model) and with CHARMM (non-polarizable model). We give the name of the module or routine, the real CPU time spent executing it, and the vector usage percentage. All routines are sorted with higher time-consumption listed first.  We can see that \textbf{Rel} has two kinds of hotspots~:

\begin{table}[t!]
\centering
\begin{tabular}{|l|d{4}|d{2}|}
\hline
  \multicolumn{1}{|c|}{\multirow{2}{*}{Module}}
& \multicolumn{1}{c|}{CPU Time}
& \multicolumn{1}{c|}{Vector} \\
& \multicolumn{1}{c|}{(s)}
& \multicolumn{1}{c|}{usage \%}\\
\hline
\hline
\multicolumn{3}{|c|}{\textbf{NUC} hotspots}\\
\multicolumn{3}{|c|}{Total CPU time~: \textbf{36.0896}~s}\\
\hline\hline
\textbf{vmlinux}      & 27.5005 & 100.00\\
libmkl\_avx512.so     &  5.7625 & 100.00\\
libmpi.so.12.0        &  2.7144 &   0.00\\
libc-2.17.so          &  0.0862 &   0.00\\
libmkl\_intel\_lp64.so&  0.0120 &   0.00\\
libiompstubs5.so      &  0.0090 &   0.00\\
libmpifort.so.12.0    &  0.0050 &   0.00\\
\hline\hline
\multicolumn{3}{|c|}{DHFR (AMOEBA, polarizable model)}\\
\multicolumn{3}{|c|}{\textbf{Computational} hotspots}\\
\multicolumn{3}{|c|}{Total CPU time~: \textbf{278.9512}~s (100 steps)}\\
\hline\hline
 tmatxb\_pme2  & 100.9210 & 0.00\\
 epolar1       &  52.7085 & 0.00\\
 ehal1         &  52.4679 & 0.00\\
 empole1       &  28.9127 & 0.00\\
 image         &  25.2141 & 0.00\\
 efld0\_direct2&  17.4910 & 0.00\\
 mlistcell$^\dag$ & 8.2543 & 0.00\\
 vlistcell$^\dag$ & 7.2870 & 0.00\\
 torque        &   2.1355 & 0.00\\
\hline\hline
\multicolumn{3}{|c|}{DHFR (CHARMM, non-polarizable model)}\\
\multicolumn{3}{|c|}{\textbf{Computational} hotspots}\\
\multicolumn{3}{|c|}{Total CPU time~: \textbf{24.7982}$
\mathbf{^*}$~s (100 steps)}\\
\hline\hline
 elj1$^*$      & 15.3259 & 0.00\\
 echarge1$^*$  &  6.7309 & 0.00\\
 image (1)     &  3.4130 & 0.00\\
 clistcell$^\dag$ & 2.8517 & 0.00\\
 image (2)$^*$ &  2.7414 & 0.00\\
\hline
\end{tabular}
\caption{Profiling of \textbf{Rel} using Intel VTune Amplifier. Simulations ran on one core and 100 steps. \textbf{MS} is DHFR with AMOEBA polarizable force field and with CHARMM force field (no polarization). Most important \textbf{NUC} and computational hostspots are shown in  separate frames. \textbf{vmlinux} is the system kernel, performing memory operations and system calls. For CHARMM calculation, image is splitted in two parts. The vectorized routines will use image(2). So, only the starred lines are counted in the total CPU time for comparison with \textbf{Vec}. The $\dag$ on some lines indicate routines that have not been vectorized in {\bf Vec}. Thus, they don't count in the total CPU time for comparison.}
\label{table:ReleaseProfUbiDHFR}
\end{table}

\label{Hotspots}
\begin{enumerate}
\item \textbf{NUC} hotspots : these are mainly due to libraries calls, system calls and memory management operations (initialization, copy, allocation and deallocation). 
\item Computational hotspots : these are mainly due to the computation of~:
\begin{itemize}
    \item the matrix-vector product operation applied at each iteration of the polarization solver ({\color{codepurple}\lv|tmatxb_pme2|}), which can be called up to 12 times at each step, depending upon the convergence method 
    \item the dipole polarization energy and forces ({\color{codepurple}\lv|epolar1|})
    \item the van der Waals energy and forces for the Halgren buffered 14–7 function ({\color{codepurple}\lv|ehal1|})
    \item the multipolar permanent electrostatic energy and forces ({\color{codepurple}\lv|empole1|})
    \item the right hand size of the polarization equation ({\color{codepurple}\lv|efld0_direct2|})
    \item the van der Waals energy and associated forces for the Lennard-Jones 6-12 function ({\color{codepurple}\lv|elj1|})
    \item the charge-charge interaction energy and associated forces  ({\color{codepurple}\lv|echarge1|})
\end{itemize}

The routines used to build neighbor lists ({\color{codepurple}\lv|vlistcell|}, {\color{codepurple}\lv|mlistcell|} and {\color{codepurple}\lv|clistcell |}) appear. Other widely used utility routines ({\color{codepurple}\lv|image|} and {\color{codepurple}\lv|torque|}) also appear.  
\end{enumerate}
In order to raise the performances of \textbf{Rel}, we need to address these hotspots. Two observations guide us~:
\begin{enumerate}
    \item \textbf{vmlinux} is taking almost as much CPU time as the multipole polarization energy and forces computation routines. This means the process makes many system calls and performs many memory operations. 
    \item the vector usage percentage is strictly $0.00$ for all the computation subroutines. This confirms that these routines only use scalar operations.
\end{enumerate}

The first observation led us to investigate the library and system calls and, first and foremost, to work on the memory management of a process running Tinker-HP in order to reduce memory operations.

The second observation forced us to rewrite the computation routines. Since using vector operations generally means using loops on arrays, the \emph{on the fly} method of computation in \textbf{Rel} must no longer be used. 

\section{Optimization strategy}
\subsection{Non-User Code hotspots}
\hspace{\parindent}\textbf{NUC} hotspots come from libraries calls, system calls (file open, read or close, MPI function calls,...) and {\color{codegreen}\lv|memset|}, {\color{codegreen}\lv|memcpy|} and calls to {\color{codegreen}\lv|malloc()|} or its derivatives that each process makes during its life.
\subsubsection{Libraries and System calls}
\paragraph{Libraries calls}
\hspace{\parindent}The vast majority of libraries calls comes from the Intel MKL library, which actually does computing work. As using it wisely can give a significant speedup, we have to provide the right  Fortran code (\textbf{UC}) to access the right vectorized functions of Intel MKL.
\paragraph{System calls}
\hspace{\parindent} When running, Tinker-HP reads a few files at the very beginning and outputs a log file which contains the simulation results and, periodically, a file containing atoms coordinates. As these input/output operations are done by the MPI--rank--0  process only,  the {\color{blue}\lv|open|}, {\color{blue}\lv|read|} and {\color{blue}\lv|close|}  system calls  do not really stress the \textsc{Unix/Linux} system, even if the simulation runs on millions of atoms.

 So, most of the system calls come from the \textsc{MPI} library and are due to two design choices in Tinker-HP~:
\begin{enumerate}
    \item As explained before, each Tinker-HP process hold a portion of the space (a domain) and maintains MPI communications with MPI processes that hold other domains nearby it, so that each process can exchange information  with the others and track atoms coming in or out of its own domain. As the other processes can run on other nodes, there can be even more time spent in system calls because of network transmissions. 
    \item A memory region is shared between all MPI processes on a computing node, using the sharing capabilities of the MPI library. The access control of this region is implemented through system calls (semaphores, {\color{codegreen}\lv|flock()|}...) to synchronize processes and guarantee non overlapping when writing data.
\end{enumerate}

Minimizing the time spent in system calls is not so easy. We tried different distributions of the MPI processes over the nodes to favour local MPI communications, but that did not give convincing results. We also tried to improve the use of MPI framework by masking communications with computations. We used non blocking versions of {\color{blue}\lv|MPI_SEND|} and {\color{blue}\lv|MPI_RECEIVE|} functions, and did some calculations before calling the corresponding {\color{blue}\lv|MPI_WAIT|}. The performance gain is not really noticeable for now. But improving this part of the code is a work in progress.
\subsubsection{Memory management}
\begin{figure}[htbp!]
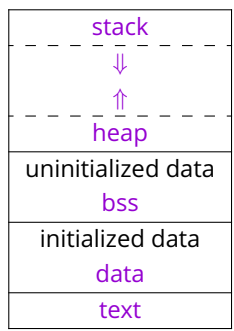

    \centering
    \begin{tabular}{|c|}
         \hline
         {\color{codepurple} stack} \\
         \hdashline
         \color{codepurple}$\Downarrow$ \\
         \color{codepurple}$\Uparrow$\\
         \hdashline
         {\color{codepurple} heap}\\
         \hline
         uninitialized data\\
         {\color{codepurple}bss}\\
         \hline
         initialized data\\
         {\color{codepurple} data}\\
         \hline
         {\color{codepurple} text}\\
         \hline
    \end{tabular}
    \caption{Memory layout of a running process. Arrows give the directions in which the zones expand.}
\label{fig:Program_layout}
\end{figure}

\hspace{\parindent}The figure~\ref{fig:Program_layout} gives a simple picture of the memory layout of a \textsc{Unix/Linux} process. The {\color{codepurple}\lv|text|} zone contains the code, the {\color{codepurple}\lv|data|} zone contains initialized data.  The {\color{codepurple}\lv|bss|} zone contains uninitialized fixed size data and has itself a fixed and limited size, set at compile time. The {\color{codepurple}\lv|heap|} zone contains allocated data and can grow or shrink, subject to allocations or de-allocations. The {\color{codepurple}\lv|stack|} zone contains a Last-In-First-Out structure, where values are pushed and pulled at each subroutine or function call or if the processor runs out of free registers.
\paragraph{\normalsize Process memory setup}
\hspace{\parindent}Historically, Tinker dynamically allocates and de-allocates all arrays it uses, because it was originally built to run on workstations with limited amount of memory and cores. These allocations are made by calls to the system {\color{codegreen}\lv|malloc()|} group of functions. As a consequence, data are put in the {\color{codepurple}\lv|heap|} section of the process, whose size is managed by the OS, allowing it to grow or shrink as needed.

As Tinker-HP is a pure MPI program which distributes data and can potentially run on hundreds of different nodes, each of them with gigabytes of memory, the problem of the memory consumption is not that important. On a computing node, each core (so, each MPI process) can easily have 2 or even 4 gigabytes of memory for its own use. 

Still the size of some arrays are proportional to the size of the systems and therefore, the \textbf{MS}-size dependent data, declared when entering a subroutine, can be very large. In a normal run, each process maintains hundreds of numbers for each atom it holds. And we can have thousands of atoms held by each process and tens of MPI processes on one node. So, allocation and de-allocation of data for each process and in each subroutine constitutes a big stress for the OS. 

Considering that the overall size of data held by one process is often under 2Gb, and that we maintain size derived constants throughout the program, we decided to remove all dynamic allocations in the vectorized routines, and declare all arrays with fixed sizes. That moves the data in the {\color{codepurple}\lv|bss|} section of the process, which is lighter than  {\color{codepurple}\lv|heap|} for the OS to handle, lowering the stress on it. 

\paragraph{\normalsize Memset and memcpy}
\hspace{\parindent}The execution cost of the {\color{codegreen}\lv|memset|} and {\color{codegreen}\lv|memcpy|} operations cannot be easily evaluated, as they come from the compiler libraries and are built upon the C library. But, because of their potential (big) effect on the performances (see table~\ref{table:ReleaseProfUbiDHFR} and discussion on page~\pageref{Hotspots}), these operations have been extensively tracked. 

Many of {\color{codegreen}\lv|memset|} come from unnecessary zeroing and have been easily removed. Some of them come from the use of intrinsic Fortran90 functions, where the compiler creates temporary storage and introduces {\color{codegreen}\lv|memcpy|} operations to work with it (for example, {\color{blue}\lv|PACK|}). We tried to remove the use of intrinsic functions as much as possible. Some of the {\color{codegreen}\lv|memset|} or {\color{codegreen}\lv|memcpy|} operations also come from the way Fortran passes arrays to subroutines or functions (as a whole array, or as a slice). We avoided these operations wherever possible.

After this optimization, the real CPU time on one core for \textbf{NUC} hotspots can be shorter by up to 10\%. But this depends a lot on the \textbf{MS} simulated and the activity of the \textsc{Unix/Linux} system outside of Tinker-HP.

\subsection{Computational hotspots}
\hspace{\parindent}The strategy we used can be developed in five guidelines : 
\begin{enumerate}
    \item \textbf{Rewrite all big internal loops}. As using vector operations means using arrays, big loops can be split in numerous short ones, each loop computing only one or a few quantities for all the involved atoms. This way, the quantities calculated can be kept in arrays and vector operations can be executed on them.
    \item \textbf{Cope with the way the compiler works on loops}. As stated in section~\ref{subsection:Considerations_vecto}, when the compiler tries to vectorize a loop, it can generate 3 execution loops~:
    \begin{itemize}
        \item a \textbf{Peeled loop} (\textbf{P}--loop), to treat array elements up to the first aligned one. 
        \item a \textbf{Kernel loop} (\textbf{K}--loop), to treat the biggest possible number of array elements with vector operations.
        \item a \textbf{Remainder loop} (\textbf{R}--loop), to treat elements that remain untreated by the previous loops.
    \end{itemize}
    As the \textbf{K}--loops are the most effective and the fastest loops, we must eliminate as much \textbf{P}--loops and \textbf{R}--loops as possible. We'll show below  what we did to achieve this goal.
    
    \item \textbf{Use vectorized mathematical operations} as much as possible. This can be difficult sometimes, because each compiler or library  implements them in its own way. For example, using the Intel Compiler, the {\color{blue}\lv|sqrt(X)|} function is not vectorized. But the power function {\color{blue}\lv|**|} is. So loops with {\color{blue}\lv|X**0.5|} have a better vectorization score than loops with {\color{blue}\lv|sqrt(X)|}. As the two functions can give slightly different numerical results, care must be taken to be sure to always have the same numerical results than the {\bf Rel}. %keep the big accuracy needed by Tinker-HP.
    \item \textbf{Limit the use of Fortran intrinsics} The use of Fortran intrinsics (SUM, PACK, etc...) is a common approach for vectorization in Molecular Dynamics package. But we found that, in many cases, the presence of intrinsics can prevent the compiler from finding the best optimization (see discussion about PACK on page~\pageref{PACK}). Intrinsics tend to protect data as much as possible. In doing so, they make frequent use of memset or memcpy operations which tends to result in unnecessary memory operations.
    \item \textbf{Have no dependency} between arrays in the loops, because the compiler will refuse to vectorize any loop where it cannot be sure that there is no dependency.
    
\end{enumerate}

With that in mind, knowing that Intel AVX-512 vector registers can hold eight 8-bytes reals or sixteen 4-bytes integers, we should have a significant improvement of the speed if the number of neighbouring atoms is big enough to fill them. That is generally the case in Tinker-HP calculations, except for very in-homogeneous \textbf{MS}.

To summarize, filling in the 512 bits registers in an efficient way and using as much vector operations as possible in a loop need :
\begin{itemize}
    \item \textbf{No dependency}, to be vectorized
    \item \textbf{Low number of arrays used}, to reduce the register pressure
    \item \textbf{Arrays as close as possible in memory}, to reduce cache miss and cost of memory operations
    \item \textbf{Data aligned on a suitable boundary}, to eliminate the \textbf{P}--loop
    \item \textbf{No subroutine calls, no un-vectorized math operations}, to get the best of the \textbf{K}--loop.
    \item \textbf{Loop count carefully chosen}, to eliminate the \textbf{R}--loop
    \item \textbf{No if-test}. If tests are mandatory (as in the selection process), they should be built in logical arrays before being used in the loop.
\end{itemize}
\subsubsection{Dependency}
\hspace{\parindent}Short loops calculate only one or a few unrelated quantities. They use the lower possible number of arrays. Thus, dependencies do not often occur. Where the non dependency cannot be automatically determined, we can easily see it and give directives to the compiler, or at worst rewrite the loop. 

\subsubsection{Data alignment}
\hspace{\parindent}Historically, in Tinker, data were located in commons, that were themselves organized with scientific development in mind. Some compilers have options to align commons. But they may be inefficient if data are not correctly organized, with memory representation in mind.

We decided to replace commons with modules, which have many advantages~:
\begin{itemize}
    \item Arrays can be efficiently aligned using directives when declared
    \item Overall readability is better, due to modularity
    \item Code can be introduced in modules, so we can group operations and write them once and for all.
\end{itemize}

In all the modules, we used an {\color{codepurple}\lv|ATTRIBUTE ALIGN::64|} directive for each array declaration. At the very beginning of this work, we used arrays like {\color{blue}\lv|pos(n,3)|} to represent, for example, the three spatial coordinates of an atom. But, sometimes, we saw that the initial alignment of the first row of {\color{blue}\lv|pos|} was not kept by the compiler for the following ones, preventing it from fully optimizing the code and forcing it to generate extra \textbf{P}--loops. All arrays are now mono-dimensional. The coordinates are represented by arrays like {\color{blue}\lv|Xpos(n)|}, {\color{blue}\lv|Ypos(n)|} and {\color{blue}\lv|Zpos(n)|}. The three coordinates are treated in the same loop, with no dependency, allowing for vectorization. 
 
\subsubsection{Data layouts in memory}
\begin{figure}[t!]
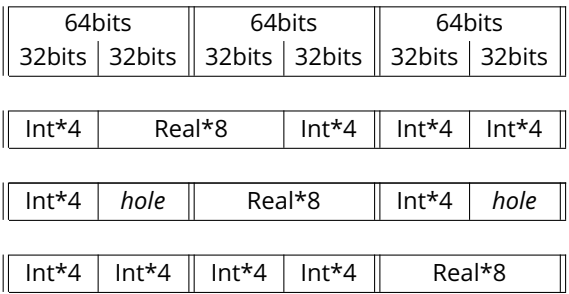

\begin{tabular}{l||c|c||c|c||c|c||}
\cline{2-7}
& \multicolumn{2}{c||}{64bits}
&\multicolumn{2}{c||}{64bits} 
&\multicolumn{2}{c||}{64bits}\\
&32bits
&32bits
&32bits
&32bits
&32bits
&32bits \\
\cline{2-7}
\multicolumn{7}{l}{}\\
\cline{2-7}
\huge\ding{172}& Int*4
&\multicolumn{2}{c|}{Real*8}
&Int*4
&Int*4 
&Int*4\\
\cline{2-7}
\multicolumn{7}{l}{}\\
\cline{2-7}
\cline{2-7}
\huge\ding{173}& Int*4
& \emph{hole}
&\multicolumn{2}{c||}{Real*8}
&Int*4
&\emph{hole} \\
\cline{2-7}
\multicolumn{7}{l}{}\\
\cline{2-7}
\huge\ding{174}& Int*4
&Int*4
&Int*4
&Int*4
&\multicolumn{2}{c||}{Real*8}\\
\cline{2-7}
\end{tabular}
\caption{Schematic picture of 3 data layouts in memory. The double vertical separators show 64 bits boundary. The single ones show 32 bits boundary.}
\label{figure:datalayout}
\end{figure}

\hspace{\parindent}The figure~\ref{figure:datalayout} shows 3 different data layouts for arrays in memory.
\begin{itemize}
    \item In the setup {\huge\ding{172}}, no {\color{codepurple}\lv|ATTRIBUTE ALIGN::64|} directive has been given. There is no memory loss, but the {\color{codegreen} real*8} array is not on a 64bits boundary. During execution, elements in this array will not be aligned. If no  {\color{codepurple}\lv|ASSUME_ALIGNED::64|} directive is given, the compiler will generate \textbf{P}-loops. If an {\color{codepurple}\lv|ASSUME_ALIGNED::64|} directive is given, no \textbf{P}-loop will be generated. But the process will pick up wrong {\color{codegreen} real*8} numbers in the \textbf{K}--loop, and give wrong results, or even crash.
    \item In the setup {\huge\ding{173}}, all arrays are aligned on a 64 bits boundary with an {\color{codepurple}\lv|ATTRIBUTE ALIGN::64|} directive. No \textbf{P}-loop  will be generated. But there can be a memory hole, if the number of elements in the integer arrays is odd. When running, the process could have to do some jumps in memory to get the {\color{codegreen} real*8} numbers, loosing time. 
    \item In the setup {\huge\ding{174}}, all arrays are aligned on 64bits  with an {\color{codepurple}\lv|ATTRIBUTE ALIGN::64|} directive. No \textbf{P}-loop  will be generated. There's no memory hole, because the number of elements in the integer arrays is kept even.
\end{itemize}
\setlength{\tabcolsep}{4pt} % Change columns separation for this tabular 
\setlength\tabcolsep\mytabsav % Restore normal columns separation

We decided to implement the setup {\huge\ding{174}}, which represents the best trade-off between speed and memory consumption. So, we ended up with the typical array declarations in a module (here, {\color{codepurple}\lv|MOD_vec_vdw|}) shown in listing~\ref{listing:declaration}, where~:
\begin{lstlisting}[label=listing:declaration,language=Fortran, firstline=2, lastline=13,caption=Typical array declarations in a module with alignment directives. Integer arrays precede real*8 arrays. Arrays are ordered as per their utilization wherever possible.]
!DIR$ ATTRIBUTES ALIGN:64::kglobvec1,kbisvec1
      integer kglobvec1(maxvlst),kbisvec1(maxvlst)
!DIR$ ATTRIBUTES ALIGN:64:: kvvec1,kvlocvec1
      integer kvvec1(maxvlst),kvlocvec1(maxvlst)
!DIR$ ATTRIBUTES ALIGN:64::rv7vec,rvvec2
      real*8 rv7vec(maxvlst),rvvec2(maxvlst)
!DIR$ ATTRIBUTES ALIGN:64::rikvec,rik2vec,rik3vec
      real*8 rikvec(maxvlst),rik2vec(maxvlst),rik3vec(maxvlst)
!DIR$ ATTRIBUTES ALIGN:64::rik4vec,rik5vec,rik6vec
      real*8 rik4vec(maxvlst),rik5vec(maxvlst),rik6vec(maxvlst)
!DIR$ ATTRIBUTES ALIGN:64::rik7vec,invrhovec,invtmpvec
      real*8 rik7vec(maxvlst),invrhovec(maxvlst),invtmpvec(maxvlst)
\end{lstlisting}

\begin{itemize}
    \item All 4-bytes integer arrays are listed before 8-bytes real ones, and each parametric dimension is a multiple of 64, eliminating the holes in the data layout and ensuring correct alignment in any cases. For example, in the listing~\ref{listing:declaration}, the parameter {\color{blue}\lv|maxvlst|}, which represents the maximum number of neighbours in van der Waals pair list,  is set to 2560.
    \item Arrays used in a loop are listed close to each others in modules and in the same order of utilization as much as possible, to reduce the time for memory operations.
\end{itemize}

For arrays declared in the subroutines (outside of modules), with shapes depending on the size of the \textbf{MS}, we calculated the next multiple of 16 bigger than the size, and used it as the dimension of the array.

Although the setup {\huge\ding{174}} implies an over-consumption of memory for arrays with \textbf{MS}--dependent sizes, this is almost not noticeable, because we add at worst 15 elements to the size, which goes usually from thousands to millions in Tinker-HP.

Almost all \textbf{P}--loops have been removed this way.
\subsubsection{Loop counts}
\hspace{\parindent}As we have many short loops in each subroutine, we really must carefully choose the loop counts. The number of sites being dependent of the size of the \textbf{MS} we work on, we cannot impose a fixed value. To be sure to completely fill the 512 bits registers at each loop, we decided to maintain two working loop counts~:
\begin{enumerate}
    \item a \textbf{Real loop} count, multiple of 8 for loops on {\color{codegreen} real*8} arrays ($8*8=64$ bytes, so 512 bits).
    \item an \textbf{Integer loop} count, multiple of 16 for loops on {\color{codegreen} integer} arrays ($16*4=64$ bytes, so 512 bits)
\end{enumerate}

As an example, if {\color{blue}\lv|nnvlst|} is the number of sites we work on, these working loop counts are computed as in listing~\ref{listing:calculationloopcount}.
\begin{lstlisting}[label=listing:calculationloopcount, language=Fortran, firstline=15, lastline=18, caption=Calculation of the working loop counts for real and integer operations]
nvloop8  = (int(nnvlst /  8) + 1) *  8
nvloop16 = (int(nnvlst / 16) + 1) * 16
nvloop8  = merge(nnvlst, nvloop8  , mod(nnvlst,  8) .eq. 0)
nvloop16 = merge(nnvlst, nvloop16 , mod(nnvlst, 16) .eq. 0)
\end{lstlisting}

Since {\color{blue}\lv|nnvlst|} can already be a multiple of 8 or 16, we should use the {\color{blue}\lv|mod|} and {\color{blue}\lv|merge|} constructs to get the smallest loop count. We used {\color{blue}\lv|nvloop8|} for {\color{codegreen}\lv|real*8|} operations, and {\color{blue}\lv|nvloop16|} for {\color{codegreen}\lv|integer|} operations. {\color{codegreen}\lv|real*8|} arrays are loaded in registers by chunks of 8 elements, and {\color{codegreen}\lv|integer|} arrays by chunks of 16 elements. We eliminated almost all the \textbf{R}--loops this way.

The flaw of this method is that, when {\color{blue}\lv|nnvlst|} is very low, we do an overwork. However, the number of neighbours is generally big enough (typically between 200 and 2000), so these extra iterations are not too expensive. Furthermore, they are executed as \textbf{K}--loops, where maximum vectorization is in effect. We just have to remember that we have now extra (useless) calculated values and find a way to drop them in our calculations.

\subsubsection{Design of loops}
\hspace{\parindent}The use of short loops allows the programmer to better understand the code. That is exactly the same for the compiler~! So, loops in \textbf{Vec} are :
\begin{itemize}
    \item \textbf{simple}. We use a small number of arrays (a maximum of 8 seems to be a good start). Due to the intrinsic complexity of the mathematical model, we use many temporary arrays.
    \item \textbf{short}. The loops contains 10 instructions at most, and 3 or 4 most of the time. 
    \item \textbf{mostly if-test free}. Most of the time, an  if-test in a loop prevents the compiler from vectorizing. For the loops in the neighbours selection process, which cannot be if-test free by nature, tests are built as logical array masks before being used. This way, loops using these masks can easily be vectorized.
\end{itemize}

\subsubsection{Editing considerations}
\hspace{\parindent}The large refactoring effort on the {\bf Rel} code may seem to be of limited efficacy at first glance. But we found that this allowed us to better understand what the code does, which was a crucial step in replacing the internal big loop with multiple short loops. It was then far easier to reason about vectorization, and modifications and debugging were simplified.

\section{Vectorized loops in Tinker-HP}
\hspace{\parindent}We have 2 kinds of vectorized loops in Tinker-HP :
\begin{itemize}
    \item \textbf{selection loops}
that select sites using various cutoffs and work on integers or logicals \item\textbf{compute loops} that compute quantities and work on reals. \end{itemize}

We will show each of them in details, and give some insights on how we have improved the vectorization. The typical loops have been extracted from {\color{codepurple}\lv|ehal1vec|}, which use the module {\color{codepurple}\lv|MOD_vec_vdw|} shown in listing~\ref{listing:declaration}.
\subsection{Typical selection loop}

\hspace{\parindent}We built a selection mask {\color{blue}\lv|mask1|} with the appropriate test, using {\color{blue}\lv|nvloop16|} here, as we work on integers (listing~\ref{listing:creationmask}). 
We first tell the compiler to assume the loop count value is a multiple of 16, so that it does not generate any \textbf{R}--loop. We really have to be sure of that, otherwise the process will pick up numbers from other arrays in the memory and wrong results or bad crashes will occur. 
\begin{lstlisting}[label=listing:creationmask, language=Fortran, firstline=20, lastline=24, caption=Loop creating a logical mask.]
!DIR$ ASSUME (MOD(nvloop16,16).eq.0)
 do k = 1, nvloop16
    mask1(k)= (kvlocvec(k) /= 0).and.(kbisvec (k) <= nbloc)
&                               .and.(kvlocvec(k) <= nbloc)
 enddo
\end{lstlisting}

 The vectorization report on listing~\ref{listing:creationreport} shows that the loop is perfectly vectorized.
\begin{lstlisting}[label=listing:creationreport, firstline=2, lastline=14, caption={Vectorization report for the mask creation. Recall that the speedup reported is not in time of execution, but in number of operations.}]
   LOOP BEGIN at (284,10)
      reference mask1(k) has aligned access
      vector length 16
      normalized vectorization overhead 0.660
      LOOP WAS VECTORIZED
      unmasked aligned unit stride loads: 3
      unmasked aligned unit stride stores: 1
      --- begin vector cost summary ---
      scalar cost: 44
      vector cost: 3.120
      estimated potential speedup: 13.840
      --- end vector cost summary ---
   LOOP END
\end{lstlisting}

\begin{lstlisting}[label=listing:selectionpack,language=Fortran,firstline=26,lastline=30,caption=First attempt with the pack function.]
         nnvlst1   = count(mask1)
         kglobvec1 = pack(kglobvec,mask1)
         kbisvec1  = pack(kbisvec ,mask1)
         kvvec1    = pack(kvvec   ,mask1)
         kvlocvec1 = pack(kvlocvec,mask1)
\end{lstlisting}

We then applied the mask on the set of atoms we worked on to select those of interest. In Fortran, we have an intrinsic function {\color{blue}\lv|PACK|} that does exactly this. So, in a first attempt, we wrote the selection loop as shown in the listing~\ref{listing:selectionpack}.\label{PACK} 

Unfortunately, because of the {\color{blue}\lv|PACK|} function, which does an implicit loop over all the elements of the array it works on, each line was seen as an independent loop by the compiler, and optimization was made on that line only. 

The corresponding Fortran compiler report on  listing~\ref{listing:report1} shows that a vectorized loop is generated for the {\color{blue}\lv|count|} line, with a vector cost of $0.810$. Then, the first {\color{blue}\lv|PACK|} line generates 3 loops :
\begin{itemize}
    \item one over the 2-bytes logical array {\color{blue}\lv|mask1|} (vector length 32) with a vector cost of $1.250$.
    \item one over the 4-bytes integer array {\color{blue}\lv|kglobvec|} (vector length 16) with a vector cost of $1.000$.
    \item one reduced to a {\color{codegreen}\lv|memset|} or {\color{codegreen}\lv|memcpy|} for the assignment of {\color{blue}\lv|kglobvec1|}
\end{itemize}

The total vector cost is $2.250$ for one {\color{blue}\lv|PACK|} operation. We also have 3 loads and 1 store for each {\color{blue}\lv|PACK|} line.

For this selection loop, we obtained a total vector cost of $0.810 + 4*2.250 = 9.810$, plus 4 {\color{codegreen}\lv|memset|} or {\color{codegreen}\lv|memcpy|}, and a total of 13 loads and 4 stores. We cannot easily know the final cost of this selection loop, because, as stated before, the implementation of the 4 memory operations depends on the compiler.
\begin{lstlisting}[label=listing:report1,firstline=16, lastline=58, caption=Vectorization report for the selection loop (pack version)]
   LOOP BEGIN at (309,22)
      vector length 16
      unroll factor set to 2
      normalized vectorization overhead 1.192
      LOOP WAS VECTORIZED
      unmasked aligned unit stride loads: 1
      --- begin vector cost summary ---
      scalar cost: 11
      vector cost: 0.810
      estimated potential speedup: 13.330
      --- end vector cost summary ---
   LOOP END

   LOOP BEGIN at (311,22)
      vector length 32
      normalized vectorization overhead 0.800
      LOOP WAS VECTORIZED
      unmasked aligned unit stride loads: 1 
      --- begin vector cost summary ---
      scalar cost: 11 
      vector cost: 1.250 
      estimated potential speedup: 8.710 
      --- end vector cost summary ---
   LOOP END

   LOOP BEGIN at (311,22)
      reference kglobvec(:) has aligned access
      vector length 16
      normalized vectorization overhead 0.188
      LOOP WAS VECTORIZED
      unmasked aligned unit stride loads: 2 
      masked unaligned unit stride stores: 1 
      --- begin vector cost summary ---
      scalar cost: 12 
      vector cost: 1.000 
      estimated potential speedup: 11.980 
      --- end vector cost summary ---
      vector compress: 1 
   LOOP END

   LOOP BEGIN at (311,10)
      loop was not vectorized: loop was transformed to memset or memcpy
   LOOP END
\end{lstlisting}
\begin{lstlisting}[label=listing:selectionnopack,language=Fortran, firstline=32, lastline=43,caption=Final selection loop with no PACK function.]
         kk = 0
!DIR$ ASSUME (MOD(nvloop16,16).eq.0)
         do k = 1, nvloop16
            if (mask1(k)) then
               kk = kk + 1
               kglobvec1(kk) = kglobvec(k)
               kbisvec1 (kk) = kbisvec (k)
               kvvec1   (kk) = kvvec   (k)
               kvlocvec1(kk) = kvlocvec(k)
            endif
         enddo
         nnvlst1 = kk
\end{lstlisting}

To get a controlled and constant vector cost, whichever compiler we use, we decided to get rid of the {\color{blue}\lv|PACK|} function. 

After all, packing data is just a matter of selecting array elements and putting them contiguously in a new array. So, we ended up with a functionally equivalent loop depicted in listing~\ref{listing:selectionnopack}.

Although there is a test in this loop, the corresponding Fortran compiler report (listing~\ref{listing:report2})  clearly shows that~:
\begin{lstlisting}[label=listing:report2, firstline=60, lastline=76, caption=Vectorization report for the selection loop.]
   LOOP BEGIN at (298,10)
      reference kglobvec(k) has aligned access
      reference kbisvec (k) has aligned access
      reference kvvec   (k) has aligned access
      reference kvlocvec(k) has aligned access
      vector length 16
      normalized vectorization overhead 0.300
      LOOP WAS VECTORIZED
      unmasked aligned unit stride loads: 5
      masked unaligned unit stride stores: 4
      --- begin vector cost summary ---
      scalar cost: 18
      vector cost: 2.500
      estimated potential speedup: 7.090
      --- end vector cost summary ---
      vector compress: 4
   LOOP END
\end{lstlisting}
\begin{itemize}
    \item The loop is vectorized
    \item Every reference is aligned, so are the loads. The stores cannot be aligned, because of the packing.
    \item The vector length is 16 which means 16 integers will be picked up in each operation
    \item The potential speedup is more than 7. This is very good in the presence of a test. 
    \item 4 vector compress instructions are generated, which correspond to the 4 assignments.
\end{itemize}

The last remark is very interesting : the Intel compiler was able to recognize this construct as a packing loop, and implemented it directly with {\color{codegreen}\lv|vpcompressd|} instructions, which are Intel AVX-512 pack instructions at the machine code level. 

A look to the assembly code in listing~\ref{listing:code1} confirms that the {\color{codegreen}\lv|vpcompressd|} instructions operate on {\color{red}\lv|zmm|} pure 512-bit vector registers.

\begin{lstlisting}[label=listing:code1, firstline=2, lastline=8, language=Fortran, caption=Typical selection loop assembly code extract showing the vpcompressd instructions.]
 vpcompressd %zmm3, vec_mp_kglobvec1_(,%rcx,4){%k1}
 vpcompressd %zmm4, vec_mp_kbisvec1_(,%rcx,4){%k1}
 vpcompressd %zmm5, vec_vdw_mp_kvvec1_(,%rcx,4){%k1}
 addl        %edx, %r14d
 incl        %edx
 movslq      %edx, %rdx
 vpcompressd %zmm6, -4+vec_vdw_mp_kvlocvec1_(,%rdx,4){%k1}
\end{lstlisting}

We obtained a vector cost of only $2.500$ and no {\color{codegreen}\lv|memset|} or {\color{codegreen}\lv|memcpy|}. That is 4 times smaller than the {\color{blue}\lv|PACK|} version, and much more if we count the {\color{codegreen}\lv|memset|} or {\color{codegreen}\lv|memcpy|} operations. The number of loads is also reduced by a factor of 3. This version is really faster than the first one.

\subsection{Typical compute loop} 
\hspace{\parindent}The calculation loops follow the scheme we have described above. They are short, simple and easy to read. The listing~\ref{listing:calculation} shows a typical compute loop. 
\begin{lstlisting}[label=listing:calculation, language=Fortran, firstline=45, lastline=55, caption={A typical compute loop. Starting from the already available rik2vec and rikvec, it computes all the powers of rikvec and intermediate quantities needed by the Halgren buffered function. Notice that there are 8 instructions and 11 different array references.}]
!DIR$ ASSUME (MOD(nvloop8,8).eq.0)
do k = 1, nvloop8
   rv7vec (k)   = rvvec2 (k)  ** 7
   rik3vec(k)   = rik2vec(k)  * rikvec(k)
   rik4vec(k)   = rik3vec(k)  * rikvec(k)
   rik5vec(k)   = rik4vec(k)  * rikvec(k)
   rik6vec(k)   = rik5vec(k)  * rikvec(k)
   rik7vec(k)   = rik6vec(k)  * rikvec(k)
   invrhovec(k) = (rik7vec(k) + ghal * rv7vec(k)) ** - one
   invtmpvec(k) = (rikvec (k) + dhal * rvvec2(k)) ** - one
enddo
\end{lstlisting}
\begin{lstlisting}[label=listing:report3, firstline=78, lastline=113, caption={Vectorization report for the compute loop. As all arrays here are aligned, no \textbf{P}--loop are generated by the compiler. Because of the loop count, no \textbf{R}--loop are generated either. }
]
   LOOP BEGIN at (417,10)
      reference rvvec2   (k) has aligned access
      reference rv7vec   (k) has aligned access
      reference rik3vec  (k) has aligned access
      reference rik2vec  (k) has aligned access
      reference rikvec   (k) has aligned access
      reference rik4vec  (k) has aligned access
      reference rik3vec  (k) has aligned access
      reference rikvec   (k) has aligned access
      reference rik5vec  (k) has aligned access
      reference rik4vec  (k) has aligned access
      reference rikvec   (k) has aligned access
      reference rik6vec  (k) has aligned access
      reference rik5vec  (k) has aligned access
      reference rikvec   (k) has aligned access
      reference rik7vec  (k) has aligned access
      reference rik6vec  (k) has aligned access
      reference rikvec   (k) has aligned access
      reference rik7vec  (k) has aligned access
      reference rv7vec   (k) has aligned access
      reference invrhovec(k) has aligned access
      reference rikvec   (k) has aligned access
      reference rvvec2   (k) has aligned access
      reference invtmpvec(k) has aligned access
      vector length 8
      normalized vectorization overhead 0.049
      LOOP WAS VECTORIZED
      unmasked aligned unit stride loads: 14 
      unmasked aligned unit stride stores: 8 
      --- begin vector cost summary ---
      scalar cost: 272 
      vector cost: 25.750 
      estimated potential speedup: 10.490 
      vectorized math library calls: 2 
      --- end vector cost summary ---
   LOOP END
\end{lstlisting}

We first tell the compiler to assume the loop count value is a multiple of 8 (we work on reals here). All arrays are independent and used in the order they were declared in the module (see listing~\ref{listing:declaration}).%All arrays are independent and used after being completely set. %So the compiler can easily optimize the loads and stores.

We can see from the corresponding Fortran compiler report in the listing~\ref{listing:report3} that~:
\begin{itemize}
    \item The loop is vectorized (no dependency).
    \item Every reference is aligned, so are the loads and stores.
    \item The vector length is 8 which means 8 numbers will be picked up in each operation
    \item The potential speedup is around $10.5$.
    \item 2 vectorized math library calls are made for the 2 {\color{blue}\lv|**|} function.
\end{itemize}
\begin{lstlisting}[label=listing:code2, firstline=10, lastline=53, caption=Typical calculation loop assembly code showing vector only operations. Loads and stores have been optimized.]
 jle       ..B2.324
 cmpl      $8, %r13d
 jl        ..B2.460
 movl      %r13d, %edx
 xorl      %eax, %eax
 andl      $-8, %edx
 movslq    %edx, %rdx
 vbroadcastsd vdwpot_mp_ghal_(%rip), %zmm17
 vbroadcastsd vdwpot_mp_dhal_(%rip), %zmm16
 vmovups   -816(%rbp), %zmm20
 movl      %r14d, -456(%rbp)
 movq      %rdx, %r14
 movq      %r12, -152(%rbp)
 movq      %rax, %r12
 # MAIN VECTOR TYPE: 64-bits floating point
 vmovups   vec_vdw_mp_rikvec_(,%r13,8), %zmm19           
 vmovups   vec_vdw_mp_rvvec2_(,%r13,8), %zmm18           
 vmulpd    vec_vdw_mp_rik2vec_(,%r13,8), %zmm19, %zmm5   
 vmulpd    %zmm18, %zmm18, %zmm2                         
 vmulpd    %zmm19, %zmm5, %zmm6                          
 vmulpd    %zmm2, %zmm2, %zmm3                           
 vmulpd    %zmm18, %zmm2, %zmm4                          
 vmovupd   %zmm5, vec_vdw_mp_rik3vec_(,%r13,8)           
 vmulpd    %zmm19, %zmm6, %zmm7                          
 vmulpd    %zmm4, %zmm3, %zmm0                           
 vmovupd   %zmm6, vec_vdw_mp_rik4vec_(,%r13,8)           
 vmulpd    %zmm19, %zmm7, %zmm8                          
 vmovupd   %zmm0, vec_vdw_mp_rv7vec_(,%r13,8)            
 vmovupd   %zmm7, vec_vdw_mp_rik5vec_(,%r13,8)           
 vmulpd    %zmm19, %zmm8, %zmm9                          
 vmovupd   %zmm8, vec_vdw_mp_rik6vec_(,%r13,8)           
 vfmadd213pd %zmm9, %zmm17, %zmm0                        
 vmovupd   %zmm9, vec_vdw_mp_rik7vec_(,%r13,8)           
 vmovaps   %zmm20, %zmm1
 call      *__svml_pown8_z0@GOTPCREL(%rip)
 vfmadd231pd %zmm16, %zmm18, %zmm19
 vmovaps   %zmm20, %zmm1
 vmovupd   %zmm0, vec_vdw_mp_invrhovec_(,%r12,8)
 vmovaps   %zmm19, %zmm0
 call      *__svml_pown8_z0@GOTPCREL(%rip)
 vmovupd   %zmm0, vec_vdw_mp_invtmpvec_(,%r12,8)
 addq      $8, %r12
 cmpq      %r14, %r12
 jb        ..B2.322
\end{lstlisting}

A look to the assembly code in listing~\ref{listing:code2} shows that all multiplications are done with vector operations {\color{codegreen}\lv|vmulpd|} and {\color{codegreen}\lv|vfmadd213pd|} and {\color{codegreen}\lv|vfmadd231pd|}, which are fused multiply-add operations. These vector instructions operate on {\color{red}\lv|zmm|} registers. We can also see the two calls to the vectorized version of the {\color{blue}\lv|**|} function so we are fully using Intel AVX-512 capabilities.

If ever we had used a division, instead of {\color{blue}\lv|**|\color{red}\lv|- one|}, we would have got :
\begin{lstlisting}[firstline=115, lastline=117, caption= Excerpt of a  vectorization report for the compute loop with division.]
      estimated potential speedup: 4.610
      divides: 2
      --- end vector cost summary ---
\end{lstlisting}
The estimated potential speedup in this case is less than half the previous one. And the utilization of the vector units is not so optimal.

So, a careful reading of the vectorization report is always necessary to ensure the best choices have been made.
\subsubsection{Final profile for \textbf{Vec}.}
\hspace{\parindent}The tables~\ref{table:VectorizedProfUbiDHFR} shows the profile and the boost factors between \textbf{Rel} and \textbf{Vec} for the final vectorized routines. 

\begin{table}[ht!]
\centering
\begin{tabular}{|l|d{4}|d{2}|d{4}|}
\hline
  \multicolumn{1}{|c|}{\multirow{2}{*}{Module}}
& \multicolumn{1}{c|}{CPU Time}
& \multicolumn{1}{c|}{Vector}
& \multicolumn{1}{c|}{Boost}\\
& \multicolumn{1}{c|}{(s)}
& \multicolumn{1}{c|}{Usage \%}
& \multicolumn{1}{c|}{factor} \\
\hline
\hline
\multicolumn{4}{|c|}{\textbf{NUC} hotspots}\\
\multicolumn{4}{|c|}{Total CPU time~: \textbf{37.3438}~s}\\
\hline\hline
\textbf{vmlinux}       & 25.4116 & 100.00&\\
libmkl\_avx512.so      &  6.1404 & 100.00&\\
libmpi.so.12.0         &  2.7094 &   0.00&\\
libmkl\_vml\_avx512.so &  2.6733 & 100.00&\\
libc-2.17.so           &  0.0703 &   0.00&\\
libmkl\_intel\_lp64.so &  0.3208 &   0.00&\\
libmpifort.so.12.0     &  0.0110 &   0.00&\\
libiompstubs5.so       &  0.0070 &   0.00&\\
\hline\hline
\multicolumn{3}{|c|}{DHFR (AMOEBA, polarizable)}&\\
\multicolumn{3}{|c|}{\textbf{Computational} hotspots} &\mathbf{2}.\mathbf{0410}\\
\multicolumn{3}{|c|}{Total CPU time~: \textbf{136.4569}~s (100 steps)}&\\
\hline\hline
 tmatxb\_pme2vec   & 62.9675 & \mathbf{ 63}.\mathbf{90} & 1.6027\\
 epolar1vec        & 29.2285 & \mathbf{ 94}.\mathbf{90} & 1.8033\\
 ehal1vec          & 19.9918 & \mathbf{ 67}.\mathbf{90} & 2.6245\\
 empole1vec        & 11.7175 & \mathbf{ 90}.\mathbf{20} & 2.4675\\
 efld0\_direct2vec &  6.9914 & \mathbf{ 82}.\mathbf{60} & 2.5018\\
 imagevec          &  4.9416 & \mathbf{100}.\mathbf{00} & 5.1024\\
 torquevec2        &  0.6186 & \mathbf{ 85}.\mathbf{70} & 3.4521\\
 \hline
 \hline
 \multicolumn{3}{|c|}{DHFR (CHARMM, no polarization)}&\\
 \multicolumn{3}{|c|}{\textbf{Computational} hotspots}& \Large\mathbf{1}.\mathbf{9535}\\
\multicolumn{3}{|c|}{Total CPU time~: \textbf{13.0355}$\mathbf{^*}$~s (100 steps)}&\\
\hline\hline
 elj1vec$^*$     & 8.2493 & \mathbf{ 64}.\mathbf{60} & 1.8578\\
 echarge1vec$^*$ & 3.6786 & \mathbf{ 90}.\mathbf{90} & 1.8297\\
 image (1)       & 3.4130 &            0.00          & 1.0000\\
 imagevec$^*$    & 1.1076 & \mathbf{100}.\mathbf{00} & 2.4751\\
 \hline 
\end{tabular}
\caption{Profiling of \textbf{Vec} using Intel VTune Amplifier. Simulations ran on one core and 100 steps. \textbf{MS} is DHFR with AMOEBA polarizable force field and with CHARMM force field (no polarization). For the vectorized routines, the Vector Usage percentages go from 63.9 to 100\%. Only the starred lines are counted in the total CPU time for DHFR with CHARMM.}
\label{table:VectorizedProfUbiDHFR}
\end{table}
\paragraph{\normalsize NUC hotspots}
\hspace{\parindent}The Real CPU Time is almost the same as for the \textbf{Rel} version. We can see a reduction of about 10\% for the real CPU time of \textbf{vmlinux}. The {\color{codepurple}\lv|libmkl_vml_avx512.so|} shared library has appeared, because we use vectorized mathematical functions and replaced all calls to the complementary error function  {\color{codegreen}\lv|erfc|}, which was in the sources of \textbf{Rel}, by calls to {\color{codegreen}\lv|vderfc|}, which is a vectorized implementation in Intel MKL library. 
\paragraph{\normalsize Computational hotspots}
\hspace{\parindent}The vector usage percentage varies between 64\% and 100\%, and the boost factors are between 1.60 and more than 5.10. The real CPU time has shrunk from $278.95s$ to $136.45s$ for AMOEBA calculation, and from $24.80s$ to $13.03s$ for CHARMM calculation, giving an overall boost factor of roughly 2.

Naive users may expect a higher boost, since Intel AVX-512 registers are designed to hold 8 reals and vector operations are built to apply on the 8 reals at the same time. But other phenomena, like indirect indexing of arrays, memory operations, vectorization coverage, and processor frequency changes are also in effect which limits the boost that can be achieved in practice. Also, even if the reported speedups for each loop can be between 6 and 15 or even more, some masked load and store operations also tend to lower the boost factor.

In a recent study, Watanabe and Nakagawa\cite{VectorizationLJ} have obtained a boost factor between 1.4 and 1.8, depending on the data setup, for the vectorization of the Lennard-Jones potential on AVX2 and AVX-512 architectures. So, achieving comparable, and often superior, boost factors on all the vectorized routines of Tinker-HP seems quite satisfactory. One interesting way to see what could be achieved is also to look at what has been done in another community code developed at Argonne and devoted to Quantum  Monte-Carlo\cite{QMCpackavx}. By definition, Monte-Carlo is the opposite of MD, as it is a highly parallel approach with few communications. In that context,  gains up to 4.5 have been obtained, highlighting some kind of upper limit of what is possible to achieve.

Watanabe and Nakagawa also showed that the boost factors are very dependent of the structure of data. This means a code written with AVX-512 capabilities in mind will not be so efficient on AVX2-capable processors. Thus, developers need to adapt their code to a specific processor in order to obtain significant boosts. But the general strategy used here should remain useful on all kind of architectures.

\section{Performance on Intel Scalable Processors}

\subsection{Sequential performance}
\hspace{\parindent}We have evaluated the overall performance boost due to Intel AVX-512 by running calculations on only one core from a dedicated node. In this situation, we can easily measure the execution time of each interesting part of the code, with limited perturbation from the MPI portions of the code or the presence of other processes.

We chose to measure three execution times :
\begin{enumerate}
    \item \textbf{time\_vdw}, which is the time taken by Van der Waals calculations. Depending on the setup, we used  {\color{codepurple} ehal1(vec)} or {\color{codepurple} elj1(vec)}.
    \item \textbf{time\_elec}, which is the time taken by electrostatic calculations. Depending on the setup, we used {\color{codepurple}echarge1(vec)} (direct charges) + reciprocal charges or {\color{codepurple}empole1(vec)} (direct multipoles) + reciprocal multipoles + {\color{codepurple} torque(vec2)}.
    \item \textbf{time\_polar}, which is the time taken by polarization calculation and by far the biggest. We used {\color{codepurple} epolar1(vec)} (direct polarization)  + {\color{codepurple}efld0\_direct(vec)} + reciprocal polarization + 9 calls to {\color{codepurple}tmatxb\_pme2(vec)} + 2 calls to {\color{codepurple}torque(vec2)}
\end{enumerate}

Notice that, as the times we measured are for the execution of a combination of vectorized subroutines and non-vectorized subroutines, they cannot be directly related to the CPU times reported in table~\ref{table:VectorizedProfUbiDHFR}.
\begin{table}[ht!]
\begin{tabular}{|l|d{4}|d{4}|d{4}|}
    \hline
    \multirow{2}{*}{\textbf{MS}}& \multicolumn{3}{c|}{Ubiquitin} \\
    \cline{2-4}
    &\multicolumn{1}{c|}{Time \textbf{Rel} (s)}
    &\multicolumn{1}{c|}{Time \textbf{Vec} (s)}
    &\multicolumn{1}{c|}{Boost}\\
    \hline
    time\_vdw   & 0.0964 & 0.0567 & 1.7002\\
    time\_elec  & 0.1352 & 0.0967 & 1.3981\\
    time\_polar & 1.2326 & 0.8758 & 1.4062\\
    \hline
    \hline
    \multirow{2}{*}{\textbf{MS}} & \multicolumn{3}{c|}{DHFR}\\
    \cline{2-4}
   &\multicolumn{1}{c|}{Time \textbf{Rel} (s)}
   &\multicolumn{1}{c|}{Time \textbf{Vec} (s)}
   &\multicolumn{1}{c|}{Boost}\\
    \hline
    time\_vdw   & 0.2359 & 0.1453 & 1.6235\\
    time\_elec  & 0.2823 & 0.2012 & 1.4031\\
    time\_polar & 2.6051 & 1.8181 & 1.4329\\
    \hline
    \hline
    \multirow{2}{*}{\textbf{MS}}&\multicolumn{3}{c|}{COX-2}\\
    \cline{2-4}
   &\multicolumn{1}{c|}{Time \textbf{Rel} (s)}
   &\multicolumn{1}{c|}{Time \textbf{Vec} (s)}
   &\multicolumn{1}{c|}{Boost}\\
    \hline
    time\_vdw   &  1.8906 &  1.1362 & 1.6639\\
    time\_elec  &  2.3816 &  1.7398 & 1.3689\\
    time\_polar & 22.2782 & 15.6152 & 1.4267\\
    \hline
    \hline
    \multirow{2}{*}{\textbf{MS}}&\multicolumn{3}{c|}{STMV}\\
    \cline{2-4}
   &\multicolumn{1}{c|}{Time \textbf{Rel} (s)}
   &\multicolumn{1}{c|}{Time \textbf{Vec} (s)}
   &\multicolumn{1}{c|}{Boost}\\
    \hline
    time\_vdw   &  1.9878 &  1.2260 & 1.6214\\
    time\_elec  &  3.8826 &  2.9314 & 1.3245\\
    time\_polar & 64.2167 & 45.4406 & 1.4132\\
   \hline
\end{tabular}

\caption{1 step measured times of execution and boost factors for different test \textbf{MS} using \textbf{Rel} or \textbf{Vec}. Simulations ran on 1 core. Values are averaged over 10 steps.}
\label{table:timings_one}
\end{table}

In this case, the boost is always a tradeoff between feeding the CPU with enough numbers, which goes better with the size of the \textbf{MS}, and minimizing the exchanges between memory and cores, which goes worse with the size. 

For every \textbf{MS}, we ran 10 calculation steps using the polarizable AMOEBA force field. We took the average value of each time, removing the smallest and the biggest. Results are given in the table~\ref{table:timings_one}.

The performance boost factors are always very good, even when the size of the \textbf{MS} is quite large (more than 1 million atoms on one core for STMV~!). The boosts we obtained are significant and justify the important vectorization efforts we made to get them.

The real performance gain should be estimated in a more \emph{realistic} situation, where Tinker-HP is running in parallel. In this case, there could be 8 to 48 processes running on one node, each competing for resources, and up to 340 nodes involved, multiplying MPI communications.
\subsection{Parallel performance}
\subsubsection{Polarizable force field~: AMOEBA }
\hspace{\parindent}We focus here on the absolute performance improvements over previous published results. As vectorization did not affect the scaling of the methods, interested readers can refer to the earlier Tinker-HP software publication for a detailed analysis of scalability \cite{Tinker-HP}.
Here, we ran calculations of 2 000 steps (4ps), with the core setups shown in table~\ref{table:MS}. The best performance was taken as the average of the 20 performance evaluations made by Tinker-HP during the run, after removing the first, middle and last values, which are lower because of the writing of intermediate files at these steps. 
\begin{table}[ht!]
\begin{tabular}{|l|d{4}|d{4}|d{4}|}
    \hline
    \multicolumn{4}{|c|}{\textbf{AMOEBA} polarizable Force Field}\\
    \hline
    \hline
     \multirow{2}{*}{\textbf{MS}}
    &\multicolumn{1}{c|}{$P_\mathbf{Rel}$}
    &\multicolumn{1}{c|}{$P_\mathbf{Vec}$}
    &\multicolumn{1}{c|}{Boost}\\
    &\multicolumn{1}{c|}{(ns/day)}
    &\multicolumn{1}{c|}{(ns/day)}
    & \multicolumn{1}{c|}{factors}\\
    \hline
    Ubiquitin       & 11.6875 & 16.9779 & 1.4526\\
    \hline
    DHFR (CPU)      &  9.1725 & 13.3312 & 1.4533\\
    DHFR (CPU2)     &  9.4761 & 14.6054 & 1.5413\\
    \hline
    Puddle          &  3.5421 &  5.2417 & 1.4798\\
    \hline
    COX-2 (CPU)     &  1.9608 &  2.9343 & 1.4965\\
    COX-2 (CPU2)    &  2.0137 &  3.1514 & 1.5650\\
    \hline
    Pond            &  1.7921 &  2.7620 & 1.5412\\
    \hline
    Lake            &  0.7066 &  1.1025 & 1.5602\\
    \hline
    STMV (CPU)      &  0.4995 &  0.7921 & 1.5858\\
    STMV (CPU2)     &  0.5152 &  0.8140 & 1.5799\\
    \hline
    Ribosome (CPU)  &  0.2295 &  0.3420 & 1.4901\\
    Ribosome (CPU2) &  0.2368 &  0.3527 & 1.4894\\
   \hline
\end{tabular}
\caption{Best production performances and boost factors for the different \textbf{MS} using \textbf{Rel} or \textbf{Vec}. For DHFR, COX-2, STMV and Ribosome, optimal results with CPU2 setup are also shown (see table~\ref{table:MS}).}
\label{table:timings_multi}
\end{table}

\begin{figure}[h]
\centering
\includegraphics[scale=0.7]{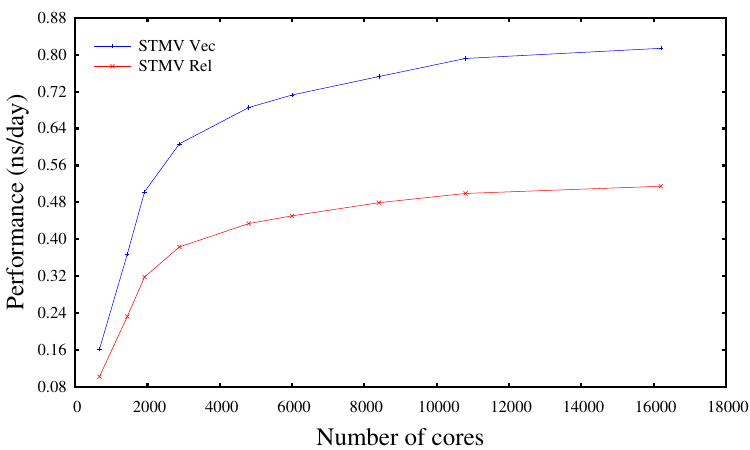}
\caption{Performance gain for the STMV using \textbf{Rel} or \textbf{Vec}. The boost factor decreases from 1.59 to 1.57 when increasing the number of cores.}
\label{fig:stmv}
\end{figure}
Results given in table~\ref{table:timings_multi} show a boost factor between 1.45 and 1.59 in parallel mode. The boost increases with the size of the \textbf{MS}, indicating a better overall utilization of the vector registers. When the \textbf{MS} is large, other phenomena (MPI memory contention, network communications, ...) result in lower boost factors. We are still able to obtain small gains with CPU2 sets, because most of the supplementary cores use vectorized routines. The results are very encouraging, especially given that not all the code has been optimized.

\begin{figure}[h]
\centering
\includegraphics[scale=0.7]{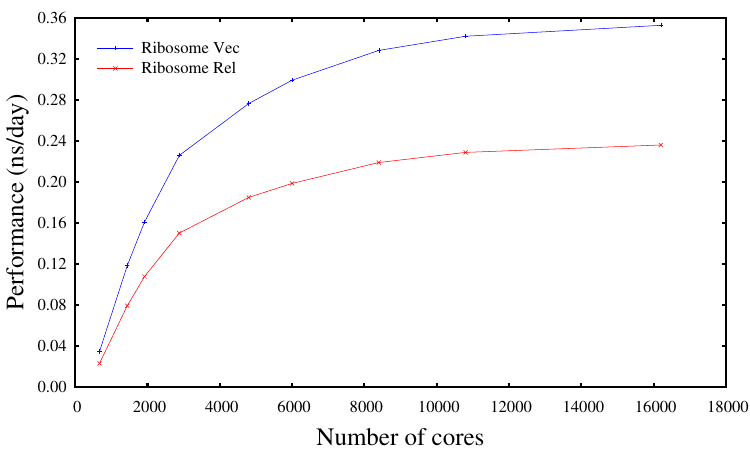}
\caption{Performance gain for the ribosome using \textbf{Rel} or \textbf{Vec}. The boost factor decreases from 1.51 to 1.49 when increasing the number of cores.}
\label{fig:ribosome}
\end{figure}

We pushed forward and tried simulations on STMV and Ribosome with up to 16200 cores (CPU2 set). Figures~\ref{fig:stmv} and \ref{fig:ribosome} show the performance obtained for \textbf{Rel} and \textbf{Vec} upon increasing the number of cores.

The boost factors remain relatively constant for these two \textbf{MS}. With very large number of cores (and very large number of nodes), both \textbf{Rel} and \textbf{Vec} speeds are bounded by MPI communication and memory operations. 
\subsubsection{Non-polarizable force field~: CHARMM}
\hspace{\parindent}Tinker-HP is not yet optimized for traditional simple partial charge force fields as no specific "modern algorithmics" are present. Indeed, our current implementation is essentially a massively parallel version of the initial Tinker code that was aimed toward performance comparisons for Steered Molecular dynamics between polarizable and non-polarizable approaches.\cite{SMD-Tinker} Also, we have tested a conservative molecular dynamics setup where bonds lengths are not rigid, reciprocal space computations are done at each timestep, etc. Such a setup was chosen in order to provide reference numbers, but the future performances can likely be accelerated substantially. At present, we need much more cores to get results comparable to those of other prominent codes\cite{GROMACS,NAMD,GENESIS,Bowers_2006}. Nevertheless, we decided to make performances measurements, firstly to get an idea of the boost that vectorization can provide in this case and, secondly, to know if we can still benefit from the scalability of the code, which is one of its greatest strengths. We used the same \textbf{MS} and the same CPU sets, limited to a maximum of 2 400 cores (i.e. as they were chosen for AMOEBA). 

\paragraph{\normalsize Vectorization boost}
\hspace{\parindent} The table~\ref{table:timings_nopol} shows the performances we obtained for \textbf{Rel} and \textbf{Vec}. 
\begin{table}[ht!]
\begin{tabular}{|l|d{4}|d{4}|d{4}|}
    \hline
    \multicolumn{4}{|c|}{\textbf{CHARMM} non polarizable Force Field}\\
    \hline
    \hline
     \multirow{2}{*}{\textbf{MS}}
    &\multicolumn{1}{c|}{$P_\mathbf{Rel}$}
    &\multicolumn{1}{c|}{$P_\mathbf{Vec}$}
    &\multicolumn{1}{c|}{Boost}\\
    &\multicolumn{1}{c|}{(ns/day)}
    &\multicolumn{1}{c|}{(ns/day)}
    & \multicolumn{1}{c|}{factors}\\
    \hline
    Ubiquitin   & 39.3068 & 48.8269 & 1.2422 \\
    \hline
    DHFR (CPU)  & 24.2333 & 31.7408 & 1.3098 \\
    DHFR (CPU2) & 26.4805 & 34.8272 & 1.3152 \\
    \hline
    Puddle      &  9.4749 & 12.8026 & 1.3512 \\
    \hline
    COX-2       &  8.1411 & 11.3459 & 1.3936 \\
    \hline
    Pond        &  5.1579 &  6.8394 & 1.3260 \\
   \hline
\end{tabular}
\caption{Best production performances and boost factors for different \textbf{MS} using \textbf{Rel} or \textbf{Vec} with CHARMM force field. For DHFR, optimal results with CPU2 setup are also shown (see table~\ref{table:MS}).}
\label{table:timings_nopol}
\end{table}

Overall, the speedup factor in using non-polarizable force fields is found to be between 3 and 4. The boost factors are lower than for AMOEBA, mainly because the vectorized part of the code which is actually executed is itself smaller. The results show the same behaviour as for AMOEBA as the size of the \textbf{MS} increases, with a peak value reached for smaller systems (around 200~000 atoms). Beyond this size, the non-vectorized code become the limiting speed factor.
\paragraph{\normalsize Scalability}

\begin{figure}[htb!]
\centering
\includegraphics[scale=0.7]{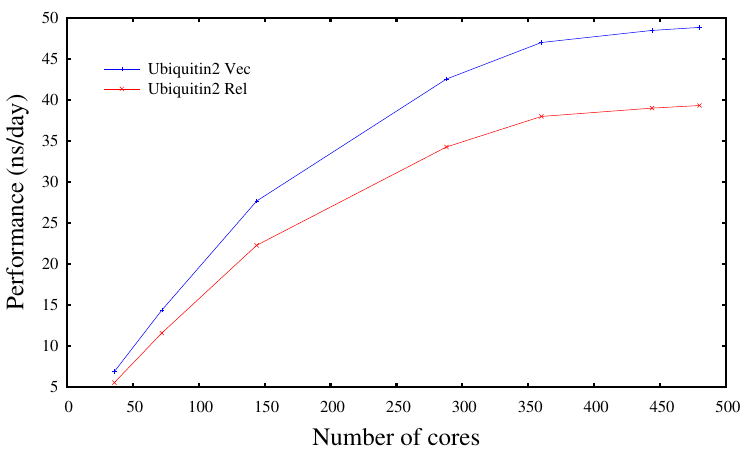}
\caption{Performance gain with CHARMM forces field for the Ubiquitin using \textbf{Rel} or \textbf{Vec}. The boost factor remains constant when increasing the number of cores.}
\label{fig:ubiquitin}
\end{figure}

\begin{figure}[htb!]
\centering
\includegraphics[scale=0.7]{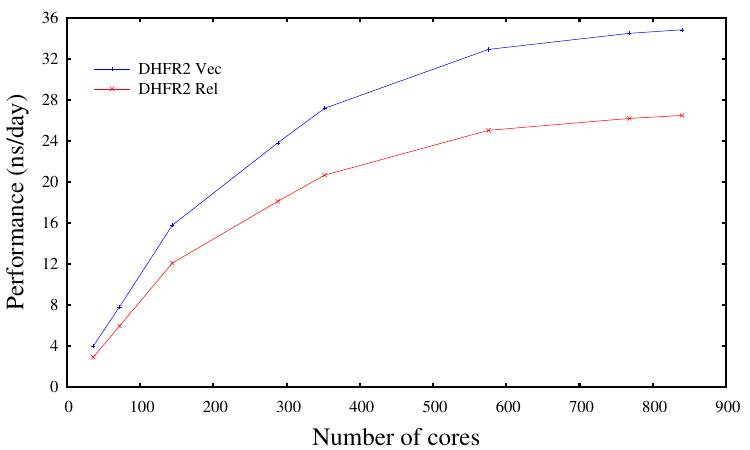}
\caption{Performance gain with CHARMM forces field for the DHFR using \textbf{Rel} or \textbf{Vec}. The boost factor remains almost constant when increasing the number of cores.}
\label{fig:dhfr}
\end{figure}

\begin{figure}[htb!]
\centering
\includegraphics[scale=0.7]{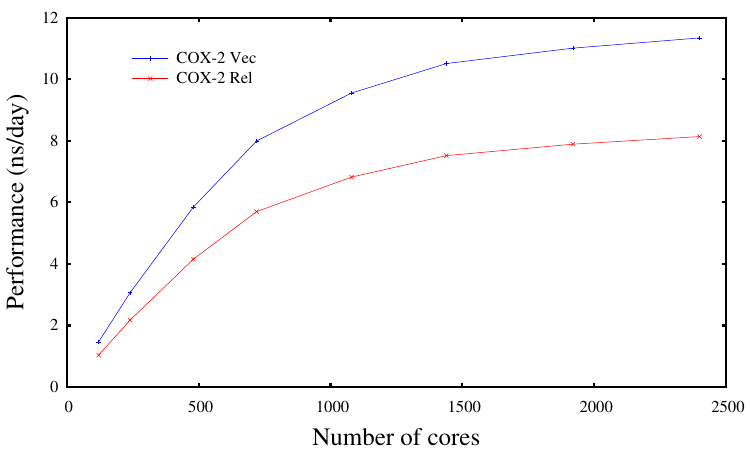}
\caption{Performance gain with CHARMM forces field for the COX-2 using \textbf{Rel} or \textbf{Vec}. The boost factor slightly decrease  when increasing the number of cores.}
\label{fig:cox2}
\end{figure}

\hspace{\parindent} We tested the scalability of the code with three \textbf{MS}~: Ubiquitin, DHFR and COX-2. As for AMOEBA, we ran for 2000 steps with increasing number of cores, and took the average performance given by the code. Figures \ref{fig:ubiquitin}, \ref{fig:dhfr} and \ref{fig:cox2} show the performance obtained for \textbf{Rel} and \textbf{Vec} when increasing the number of cores.

For all \textbf{MS} simulated, the scalability is still very good. The boost factor remains almost constant for the two smaller \textbf{MS} (Ubiquitin and DHFR). For COX-2, the boost factor decreases from 1.41 to 1.39 when increasing the number of cores because, with 2~400 cores, communications tends to lower the benefit of the vectorization. In practice, this version of the code is a first step towards an efficient engine for non-polarizable MD but work is still required and is in progress to obtain better performance with updated code.

\subsection{Perspectives on Tinker-HP 1.2 performance} \label{subsection:Perspectives}
\hspace{\parindent}This section gives an indication of the performance gains that will appear in the forthcoming Tinker-HP Release 1.2 version (\textbf{Rel2}). Indeed, despite being not fully vectorized, this major update proposes significant algorithmic speedups.  For now, we can point out that a strong performance gain without accuracy loss is observed in using Tinker-HP with the new multi-timestep BAOAB-RESPA1 integrator\cite{baoab-respa1}, and with hydrogen mass repartitioning.
This newly introduced integrator splits the energy terms in three levels evaluated at different timesteps: the bonded terms are evaluated every 1 fs, the non-bonded terms (including polarization) are split into short and long range, the short-range being evaluated every 10/3 fs and the long range every 10 fs. Furthermore, short-range polarization is treated with the non-iterative TCG-1 (Truncated Conjugate Gradient) solver\cite{aviatjctc2017,aviatjpc2017} and the outer-level uses the Divide-and-Conquer Jacobi Iterations (DC-JI) \cite{Nocito2018} approach, offering a net global acceleration a factor of 4.91 compared to standard 1 fs/Beeman/ASPC (7 without ASPC) simulations without loss of accuracy, enabling an accurate evaluation of properties such as free energies\cite{baoab-respa1}.

Preliminary results (where not all routines are yet vectorized) are reported in Table~\ref{table:timings_multi2}. We intend to review the full 1.2 vectorized version Tinker-HP in a future update of this living review.

\begin{table}[ht!]
\begin{tabular}{|l|d{4}|d{4}|d{4}|}
    \hline
    \multicolumn{4}{|c|}{\textbf{AMOEBA} polarizable Force Field}\\
    \hline
    \hline
     \multirow{2}{*}{\textbf{MS}}
    &\multicolumn{1}{c|}{$P_\mathbf{Rel2}$  }
    &\multicolumn{1}{c|}{$P_\mathbf{Rel2-multi}$ }
    &\multicolumn{1}{c|}{$P_\mathbf{Vec2-multi}$}\\
    &\multicolumn{1}{c|}{(ns/day)}
    &\multicolumn{1}{c|}{(ns/day)}
    &\multicolumn{1}{c|}{(ns/day)}\\
    \hline
    Ubiquitin       & 11.6875 & 28.28  & 40.32\\
    \hline
    DHFR (CPU)      &  9.1725 & 22.20  & 32.26\\
    DHFR (CPU2)     &  9.4761 & 22.93  & 35.33\\
    \hline
    Puddle          &  3.5421 &  8.57  & 12.68\\
    \hline
    COX-2 (CPU)     &  1.9608 &  4.74  &  7.09\\
    COX-2 (CPU2)    &  2.0137 &  4.87  &  7.65\\
    \hline
    Pond            &  1.7921 &  4.34  &  6.69\\
    \hline
    Lake            &  0.7066 &  1.70  &  2.65\\
    \hline
    STMV (CPU)      &  0.4995 &  1.21  &  1.92\\
    STMV (CPU2)     &  0.5152 &  1.25  &  1.97\\
    \hline
    Ribosome (CPU)  &  0.2295 &  0.55  &  0.82\\
    Ribosome (CPU2) &  0.2368 &  0.57  &  0.85\\
   \hline
\end{tabular}
\caption{Best production performances  for the different \textbf{MS} using \textbf{Rel2}, \textbf{Rel2-multi} (multi-timestep) and \textbf{Vec2-multi} (multi-timestep). For DHFR, COX-2, STMV and Ribosome, optimal results with CPU2 setup are also shown (see table~\ref{table:MS}).}
\label{table:timings_multi2}
\end{table}

As of July 2019, we have vectorized the neighbor list building routines and made improvements to the vectorization of other routines. The  table~\mbox{\ref{table:VectorizedProfUbiDHFRUpdate}} shows the new boost factors for the computational hotspots. The individual speedups for neighbor list subroutines range from $1.54$ to $2.73$, raising the overall vectorization boost factor to $2.45$ for the polarizable force field and $2.92$ for classical force fields. The overall performance boost (parallel gain) for classical forces field thus increases from 1.4 to 2.0 bringing interesting perspectives towards a future highly optimized classical force field MD engine.
\begin{table}[ht!]
\centering
\begin{tabular}{|l|d{4}|d{2}|d{4}|}
\hline
  \multicolumn{1}{|c|}{\multirow{2}{*}{Module}}
& \multicolumn{1}{c|}{CPU Time}
& \multicolumn{1}{c|}{Vector}
& \multicolumn{1}{c|}{Boost}\\
& \multicolumn{1}{c|}{(s)}
& \multicolumn{1}{c|}{Usage \%}
& \multicolumn{1}{c|}{factor} \\
\hline
\hline
\multicolumn{3}{|c|}{DHFR (AMOEBA, polarizable)}&\\
\multicolumn{3}{|c|}{\textbf{Computational} hotspots} &\mathbf{2}.\mathbf{4566}\\
\multicolumn{3}{|c|}{Total CPU time~: \textbf{119.8797}~s (100 steps)}&\\
\hline\hline
 tmatxb\_pme2vec   & 57.4947 & \mathbf{100}.\mathbf{00} & 1.7553\\
 epolar1vec        & 21.0042 & \mathbf{100}.\mathbf{00} & 2.5094\\
 ehal1vec          & 16.6440 & \mathbf{ 78}.\mathbf{10} & 3.1524\\
 empole1vec        &  7.8784 & \mathbf{ 90}.\mathbf{20} & 3.6699\\
 efld0\_direct2vec &  6.4751 & \mathbf{100}.\mathbf{00} & 2.7013\\
 vlistcellvec      &  4.7310 & \mathbf{100}.\mathbf{00} & 1.5403\\
 mlistcellvec      &  3.0972 & \mathbf{100}.\mathbf{00} & 2.6651\\
 imagevec          &  2.5108 & \mathbf{100}.\mathbf{00} &10.0422\\
 torquevec2        &  0.4432 & \mathbf{100}.\mathbf{00} & 4.8183\\
 \hline
 \hline
 \multicolumn{3}{|c|}{DHFR (CHARMM, no polarization)}&\\
 \multicolumn{3}{|c|}{\textbf{Computational} hotspots}& \Large\mathbf{2}.\mathbf{9201}\\
\multicolumn{3}{|c|}{Total CPU time~: \textbf{10.6051}~s (100 steps)}&\\
\hline\hline
 elj1vec         & 6.3047 & \mathbf{ 75}.\mathbf{00} & 2.4309\\
 echarge1vec     & 2.6411 & \mathbf{100}.\mathbf{00} & 2.5486\\
 clistcellvec    & 1.0424 & \mathbf{100}.\mathbf{00} & 2.7357\\
 imagevec        & 0.6169 & \mathbf{100}.\mathbf{00} & 9.9763\\
 \hline 
\end{tabular}
\caption{Profiling of \textbf{Vec} using Intel VTune Amplifier.  Simulations ran on one core and 100 steps. \textbf{MS} is DHFR with AMOEBA polarizable force field and with CHARMM force field (no polarization). For the vectorized routines, the Vector Usage percentages go from 78.0 to 100\%. As imagevec has been fully vectorized, there is no more separation for the image CPU time in the CHARMM part of the table. Neighbor list building routines have been fully vectorized. So, the lines with $\dag$ in the table~\ref{table:ReleaseProfUbiDHFR} have been reintegrated in the total CPU time to compute the general boost factor.}
\label{table:VectorizedProfUbiDHFRUpdate}
\end{table}

Finally, beside the focus on the AMOEBA polarizable force field, performances will be given for other polarizable models as well as on classical force fields (CHARMM, AMBER, OPLS etc...). For now, despite the non-optimization and the absence of use of lower precision of this part of the code, more than a 4-time speedup of the values reported in Table~\ref{table:timings_multi2} give an initial idea of the reasonable code performances for non-polarizable simulations.
\section{Conclusion}
\hspace{\parindent}In many ways this work represents a fundamental step in the evolution of the Tinker-HP software.

First, it demonstrates that new HPC architectures can offer significant acceleration to an existing massively parallel code like Tinker-HP. A brute  performance boost factor between 1.32 and 1.70 can be achieved on computationally intensive parts of the code, leading to an overall acceleration factor between 1.45 and 1.59 for AMOEBA (1.24 and 1.40 for CHARMM) under realistic conditions, including the simulation of  molecular systems with millions of atoms. Considering that the many current calculations require a total simulation time of a several microseconds, such a speed gains represent major progress.

Second, it shows that improved speed is not just available from raising the frequency of the CPU or buying more powerful computers. Large accelerations such as those reported here involve a close cooperation between the computational chemists, who write the code, and HPC specialists, who know how the CPU and the system software work. To get these gains, we had to dig into the pieces of code that were the most CPU consuming and to rewrite them almost completely, with simplicity and efficiency in mind. But it was worth the effort. Furthermore, considering the trends observed with prior CPUs, we anticipate that vectorization will also play an important role in future architectures.

Third, this work gives us a strategy and some methods to further improve the code. It can serve as a solid starting point for the future. We are now able to more easily adapt Tinker-HP to new underlying hardware or software advances. That will allow us to make the best of new technologies.

Of course, optimization is far from finished as some parts of the code are not yet vectorized (for example the reciprocal space computations involved in permanent electrostatics and polarization), and other sources of possible speedups exist and will be investigated. In particular, we have to review how we can improve the creation of neighbour lists, implement faster indexing of all the atoms (sorting indexes could be a solution) and achieve better masked MPI communications within computations. Decreasing precision is also possible in specific cases to gain performances while retaining sufficient accuracy. This paper will continue to be updated as we accumulate new data on Github, until a new version of this living document is pushed to review. The next iteration of the paper will also incorporate results on next-generation of Intel Xeon Scalable processors (codenamed Cascade Lake), and attempt to evolve towards an adaptation of the Tinker-HP code to any future architectures proposed by Intel. Future work will focus on the algorithmic boosting of our initial implementation of classical non-polarizable force fields. In addition, the next iteration of the paper will propose more detailed benchmarks for new polarizable approaches, including SIBFA \cite{gresh2007anisotropic,SIBFAjpp} and ongoing modifications of AMOEBA such as AMOEBA+\cite{AMOEBA+} and HIPPO\cite{electrohippo,hippoexchange}. 
\section{Acknowledgments}
%%%%%%%%%%%%%%%
% You should include all people who have filed issues that were
% accepted into the paper, or that upon discussion altered what was in the paper.
% Multiple significant contributions might mean that the contributor
% should be moved to authorship at the discretion of the authors
%
% See the policies ``Policies on Authorship'' section of https://livecoms.github.io for
% more information on deciding on authorship and author order.
%%%%%%%%%%%%%%%

\hspace{\parindent}We want to give special thanks to David Guibert and Cedric Bourasset (Center for Excellence in Parallel Programming, Atos-Bull) for initial discussions.

LHJ wishes to thank Prof Esma\"\i l Alikhani (Sorbonne Université) for sharing his 2x18 cores Skylake computer in the early phase of this work. 

LHJ also wants to thank Bernard Voillequin for fruitful discussions at the very beginning of this work. B.~Voillequin is not a scientist, and not even an engineer. But he's a man of good sense and good will. That is invaluable.

% We suggest you preserve this comment:

\section{Funding Information}
%%%%%%%
% Authors should acknowledge funding sources here. Reference specific grants.
%%%%%%%
\hspace{\parindent}This work has received funding from the European Research Council (ERC) under the European Union's Horizon 2020 research and innovation programme (grant agreement No 810367), project EMC2 (JPP). JWP and PR acknowledge support from the U.S. National Institutes of Health (R01GM106137 and R01GM114237). The authors acknowledge the generous support provided by GENCI (Grand Equipement National de Calcul Intensif, France) through the Grand Challenge and grant no A0050707671, which provided computing allocations on the Irène Joliot-Curie (TGCC, Bruyère le Châtel, France), Occigen and Frioul machines (CINES, Montpellier, France). The authors also thank the TACC (Texas Advanced Computer Center) for computational resources provided on the Stampede-2 machine.

\section*{Disclaimers}
\hspace{\parindent} Software and workloads used in performance tests may have been optimized for performance only on Intel microprocessors.

Performance tests, such as SYSmark and MobileMark, are measured using specific computer systems, components, software, operations and functions. Any change to any of those factors may cause the results to vary. You should consult other information and performance tests to assist you in fully evaluating any contemplated purchases, including the performance of that product when combined with other products.

For more information go to www.intel.com/benchmarks.

Performance results are based on testing as of March 14th 2019 and may not reflect all publicly available security updates.  See configurations described in the paper for details.  No product or component can be absolutely secure.

Intel technologies’ features and benefits depend on system configuration and may require enabled hardware, software or service activation. Performance varies depending on system configuration. Check with your system manufacturer or retailer or learn more at intel.com.

Intel, the Intel logo, Xeon, VTune and Xeon Phi are trademarks of Intel Corporation or its subsidiaries in the U.S. and/or other countries.

*Other names and brands may be claimed as the property of others. 

\section*{Author Information}
\makeorcid
\begin{tabular}{lcl}
 Luc-Henri Jolly &:& 0000-0003-3220-5759 \\
 Louis Lagardère &:& 0000-0002-7251-0910\\
 Jay W. Ponder& :  & 0000-0001-5450-9230\\
 Pengyu Ren &: &0000-0002-5613-1910\\
 Jean-Philip Piquemal&:& 0000-0001-6615-9426\\
 \end{tabular}
\bibliographystyle{vancouver-compatible}
\bibliography{Tinkerperf}

%%%%%%%%%%%%%%%%%%%%%%%%%%%%%%%%%%%%%%%%%%%%%%%%%%%%%%%%%%%%
%%% APPENDICES
%%%%%%%%%%%%%%%%%%%%%%%%%%%%%%%%%%%%%%%%%%%%%%%%%%%%%%%%%%%%

%\appendix
\onecolumn
\listoffigures
\listoftables
\lstlistoflistings
\end{document}